\documentclass[journal,twoside,web]{ieeecolor}
\usepackage{generic}
\usepackage{cite}
\usepackage{amsmath,amssymb,amsfonts}
\usepackage{algorithmic}
\usepackage{graphicx}
\usepackage{algorithm,algorithmic}
\usepackage{hyperref}
\hypersetup{hidelinks=true}
\usepackage{textcomp}
\usepackage{tikz}
\usepackage{newfloat}
\DeclareFloatingEnvironment{float}
\usepackage{pgfplots}
\pgfplotsset{compat=newest}
\newlength\hohe
\newlength\breite

\newcommand{\regret}{\mathcal R_T}
\newcommand{\norm}[1]{\left\lVert#1\right\rVert}
\newcommand{\refeq}[1]{\overset{#1}{=}}
\newcommand{\refleq}[1]{\overset{#1}{\leq}}

\newcommand{\refin}[1]{\overset{#1}{\in}}

\newcommand{\overbar}[1]{\mkern 1.5mu\overline{\mkern-1.5mu#1\mkern-1.5mu}\mkern 1.5mu}

\newcommand{\predx}{\hat{x}^{\mu}}
\newcommand{\predu}{\hat{u}^{\mu}}
\newcommand{\predz}{\hat{z}}
\newcommand{\estx}{\hat{\theta}}
\newcommand{\estu}{\hat{\eta}}
\newcommand{\estz}{\hat{\zeta}}
\newcommand{\ssinput}{\hat{u}^s}
\newcommand{\ssstate}{\hat{x}^s}
\newcommand{\ssz}{\hat{z}^s}
\newcommand{\constraintX}{\mathcal{X}}
\newcommand{\constraintU}{\mathcal{U}}
\newcommand{\constraintZ}{\mathcal{Z}}
\newcommand{\setW}{\overbar{\mathcal{W}}}
\newcommand{\setV}{\overbar{\mathcal{V}}}
\newcommand{\feasibleSS}{\mathcal{S}}
\newcommand{\stronglyfeasibleSS}{\bar{\mathcal{S}}}
\newcommand{\feasiblesequence}{\mathcal{Z}_U^\mu}
\newcommand{\measx}{\tilde{x}}
\newcommand{\RPI}{\mathcal P}
\newcommand{\noise}{\bar{w}}
\newcommand{\ball}{\mathbb{B}}
\newcommand{\controlMatrix}{S_c}

\DeclareMathAlphabet{\mymathbb}{U}{BOONDOX-ds}{m}{n}
\newcommand{\zero}{\ensuremath{\mymathbb{0}}}
\newcommand{\one}{\ensuremath{\mymathbb{1}}}

\newtheorem{defn}{Definition}
\newtheorem{thm}{Theorem}
\newtheorem{ass}{Assumption}
\newtheorem{lem}{Lemma}
\newtheorem{rem}{Remark}

\newcommand*\circled[1]{\tikz[baseline=(char.base)]{\node[shape=circle,draw,inner sep=.25pt] (char) {#1};}}

\newcommand\copyrighttext{%
	\footnotesize \textcopyright2025 IEEE. Personal use of this material is permitted.  Permission from IEEE must be obtained for all other uses, in any current or future media, including reprinting/republishing this material for advertising or promotional purposes, creating new collective works, for resale or redistribution to servers or lists, or reuse of any copyrighted component of this work in other works.}
\newcommand\copyrightnotice{%
	\begin{tikzpicture}[remember picture,overlay]
		\node[anchor=south,yshift=5pt,xshift=0pt,fill=white] at (current page.south) {\fbox{\parbox{\dimexpr.9\textwidth-\fboxsep-\fboxrule\relax}{\copyrighttext}}};
	\end{tikzpicture}%
}

\def\BibTeX{{\rm B\kern-.05em{\sc i\kern-.025em b}\kern-.08em
    T\kern-.1667em\lower.7ex\hbox{E}\kern-.125emX}}
\begin{document}	
\title{Online convex optimization for robust control of constrained dynamical systems}
\author{Marko Nonhoff, \IEEEmembership{Member, IEEE}, Emiliano Dall'Anese, \IEEEmembership{Senior Member, IEEE}, \linebreak and Matthias A. Müller, \IEEEmembership{Senior Member, IEEE}
\thanks{
This work was supported by the Deutsche Forschungsgemeinschaft (DFG, German Research Foundation) - 505182457.}
\thanks{Marko Nonhoff was supported by the 'Graduiertenakademie' of the Leibniz University Hannover.}
\thanks{The work of E. Dall'Anese was supported in part by the National Science Foundation award - 2444163.}
\thanks{Marko Nonhoff and Matthias A. Müller are with the Institute of Automatic Control, Leibniz University Hannover, Hannover, Germany (email: \{nonhoff,mueller\}@irt.uni-hannover.de). }
\thanks{Emiliano Dall'Anese is with the Department of Electrical and Computer Engineering, Boston University, Boston, MA 02215, USA (e-mail: edallane@bu.edu).}
}

\maketitle

\begin{abstract}
This article investigates the problem of controlling linear time-invariant systems subject to time-varying and a priori unknown cost functions, state and input constraints, and exogenous disturbances. We combine the online convex optimization framework with tools from robust model predictive control to propose an algorithm that is able to guarantee robust constraint satisfaction. The performance of the closed loop emerging from application of our framework is studied in terms of its dynamic regret, which is proven to be bounded linearly by the variation of the cost functions and the magnitude of the disturbances. We corroborate our theoretical findings and illustrate implementational aspects of the proposed algorithm by a numerical case study on a tracking control problem of an autonomous vehicle.
\end{abstract}


\section{Introduction} \label{sec:intro}
\copyrightnotice
In recent years, the online convex optimization (OCO) framework has emerged as a powerful approach to controller design for dynamical systems. Compared to classical numerical optimization, in OCO the cost functions are allowed to be time-varying and a priori unknown, see, e.g., \cite{Shalev-Shwartz2012, Hazan2016} for an overview. Such time-varying cost functions arise in a range of relevant applications, for example due to renewable energy generation and a priori unknown consumption in energy grids \cite{Tang2017} or in tracking control, when the desired trajectory is computed online itself \cite{Zheng2020}. Therefore, various algorithms for control of dynamical systems based on the OCO framework have recently been proposed in the literature, see, e.g., \cite{Li2019,Agarwal2019,Shi2020,Hazan2022,Nonhoff2022a,Lin2023,Karapetyan2023} and the references therein. These algorithms typically aim to track the optimal steady states of the system, which are a priori unknown and time-varying due to their dependence on the cost functions as well. The performance of the closed loop emerging from application of these algorithms is analyzed by bounding the dynamic regret, a performance measure adapted from the OCO framework. Dynamic regret $\regret$ is defined as the cumulative performance difference over an arbitrary finite horizon $T$ between the closed-loop trajectory $\{x_t,u_t\}_{t=0}^T$ and some appropriately defined benchmark $\{\chi_t,\nu_t\}_{t=0}^T$, i.e.,
\begin{equation}
	\regret := \sum_{t=0}^T \left( L_t(x_t,u_t) - L_t(\chi_t,\nu_t) \right), \label{eq:def_regret_intro}
\end{equation}
where $L_t:\mathbb{R}^n\times\mathbb{R}^m\mapsto\mathbb{R}$ is a time-varying performance measure. Recently, dynamic regret has found applications in the control literature independent of the OCO framework \cite{Didier2022,Goel2021,Martin2022,Martin2023,Zhou23CDC,Gharbi2021}, fundamental limits for the optimal achievable regret have been derived \cite{Li2019}, and its implications on the more classical notion of stability have been studied \cite{Nonhoff2023a,Karapetyan2022}.

Despite their inherent ability to operate in dynamic environments, characterized by, e.g., time-varying and a priori unknown cost functions or disturbances, the main advantages of OCO-based controllers are their low computational complexity and their ability to cope with constraints on the control input and the state of the controlled system. Such constraints are ubiquitous in real-world applications, emerging due to, e.g., actuator limitations, safety considerations, and physical limitations of the system under control. In these applications, safety guarantees in terms of constraint satisfaction are of paramount importance. In recent years, first results on OCO-based control of dynamical systems guaranteeing satisfaction of state and input constraints have been reported \cite{Nonhoff2021,Li2021,Nonhoff2024,Zhou23,Nonhoff2025}. 

A closely related line of research is so-called feedback optimization. Therein, optimization algorithms are directly employed as feedback controllers in order to steer the system under control to the solution of a (possibly time-varying) optimization problem, see, e.g., \cite{Simonetto2020} and the references therein. Typically, stability of the optimal steady state of the optimization problem is guaranteed instead of a bound on the dynamic regret \cite{Menta2018,Colombino2020,Cothren2022,Lawrence2021,Bianchin2022}. However, in the feedback optimization setting constraints are typically only considered for the optimal steady state, while pointwise in time constraints on the state of the controlled system can generally not be satisfied.

In this work, we propose a framework for robust control of dynamical systems subject to a priori unknown and time-varying cost functions, and state and input constraints that have to be met at each time instance. In particular, we consider disturbances acting on the system as well as measurement noise, which can capture, e.g., model mismatch, exogenous (uncontrollable) inputs to the system, sensor inaccuracies, state estimation error due to the application of an observer or perception-based techniques \cite{Dean2021,Cothren2022,Marchi2022}, and pseudo-measurement in the context of power systems \cite{Schenato2014,Picallo2018}. The combination of these types of disturbances with state and input constraints that have to be satisfied at all times has - to the best of the authors' knowledge - not been studied within the OCO framework, with the notable exceptions of \cite{Li2021,Zhou23,Nonhoff2024}. However, both \cite{Li2021} and \cite{Zhou23} only consider disturbances (but no measurement noise). Furthermore, \cite{Li2021} studies the problem of disturbance rejection, i.e., develops an algorithm that counteracts the disturbances and aims to drive the system to the origin. Finally, the algorithm presented in \cite{Nonhoff2024} is computationally and conceptually simple, but can be difficult to tune for a priori unknown cost functions, potentially leading to poor closed-loop performance. In contrast, we take both disturbances and measurement noise into account in this work, develop an algorithm that guarantees robust constraint satisfaction, and provide practical guidelines to choose the hyperparameters of the proposed algorithm. To this end, we apply a suitable constraint tightening using techniques from robust model predictive control (MPC) \cite{Rawlings2017,Chisci2001}. Moreover, we prove that the dynamic regret of our algorithm is bounded linearly in terms of the variation of the cost functions and the magnitude of the disturbances.

We close this section by noting the preliminary conference version containing parts of this paper \cite{Nonhoff2021}. We significantly improve the results presented therein in multiple directions. First, we consider dynamical systems with disturbances as well as measurement noise, and guarantee robust constraint satisfaction despite the presence of these uncertainties. Second, we relax restrictive assumptions, thereby improving the applicability of the proposed approach. In particular, we relax \cite[Assumption~5]{Nonhoff2021}, and allow \textit{economic} cost functions (i.e., cost functions that are not necessarily positive definite with respect to any steady state of the controlled system) by leveraging techniques from \cite{Nonhoff2022a}. To achieve the former, we suitably adapt the proposed algorithm and develop new proof techniques to ensure a sufficient rate of convergence, which is necessary to prove bounded dynamic regret. Finally, we provide a detailed numerical case study to demonstrate the applicability of the proposed algorithm in this work.

This paper is organized as follows. Section~\ref{sec:setting} formalizes the setting considered in this work. Section~\ref{sec:algo} introduces the proposed algorithm, and theoretical guarantees on constraint satisfaction and boundedness of its dynamic regret are established in Section~\ref{sec:results}. Section~\ref{sec:sim} illustrates implementational aspects of the proposed algorithm on a numerical simulation of a traffic scenario. Finally, Section~\ref{sec:conclusion} summarizes the contributions and explores directions for future research.

\textit{Notation:} The set of natural numbers (including $0$) and real numbers are $\mathbb{N}$ and $\mathbb{R}$, respectively. The set of all integers in the interval $[a,b]$, $b\geq a$, $a,b\in\mathbb R$, and the set of all integers greater than or equal to $a$ are given by $\mathbb{N}_{[a,b]}$ and $\mathbb{N}_{\geq a}$. We write the identity matrix of size $n$ and the matrix of all zeros as $I_n\in\mathbb R^{n\times n}$ and $\zero_{m,n}\in\mathbb R^{m\times n}$, where we omit the subscripts when dimensions are clear from context. For a vector $x\in\mathbb{R}^n$, $\norm{x}$ is the Euclidean norm, and for a matrix $A\in\mathbb{R}^{n\times m}$, $\norm{A}$ is the induced matrix $2$-norm. For two sets $\mathcal A,~\mathcal B\subseteq \mathbb R^n$, we denote the (relative) interior by $\text{int }\mathcal A$ ($\text{rel\,int }\mathcal A$), and Minkowski set addition and Pontryagin set difference by $\mathcal A \oplus \mathcal B := \{a+b: a\in\mathcal A,~b\in\mathcal B\}$ and $\mathcal A \ominus \mathcal B := \{a: \{a\}\oplus\mathcal B\subseteq A\}$. The diameter and radius of a set $\mathcal A$ are $d_{\mathcal A}:=\max_{a,b\in\mathcal A} \norm{a-b}$ and $r_{\mathcal A} := \max_{a\in\mathcal A} \norm{a}$. Projection of a point $x\in\mathbb R^n$ onto a convex and compact set $\mathcal A\subseteq\mathbb R^n$ is defined by $\Pi_{\mathcal A}(x) := \min_{y\in\mathcal A} \norm{x-y}^2$. For a vector $a = \begin{bmatrix} a_1^\top&\dots&a_n^\top \end{bmatrix}^\top \in \mathbb R^{nm}$, $a_i\in\mathbb R^m$ for all $i\in\mathbb{N}_{[1,n]}$, we define the block shift operator $\sigma a := \begin{bmatrix} a_2^\top & \dots & a_{n}^\top \end{bmatrix}^\top$ and the matrix that extracts the $i$-th component $T_i := \begin{bmatrix} \zero_{m,(i-1)m} & I_m & \zero_{m,(n-i)m} \end{bmatrix}$.

\section{Setting} \label{sec:setting}

We consider constrained linear time-invariant (LTI) systems
\begin{subequations} \label{eq:sys}
	\begin{align} 
		x_{t+1} &= Ax_t + Bu_t + w_t \label{eq:sys_dyn} \\
		\measx_t &= x_t + v_t \label{eq:meas} \\
		x_t &\in \constraintX,~u_t\in\constraintU \label{eq:constraints}
	\end{align}
\end{subequations}
where $t\in\mathbb{N}$, $x_t\in\mathbb R^n$ is the (real) system state, $\measx_t\in\mathbb R^n$ is the measured system state, $u_t\in\mathbb R^m$ is the control input, $w_t\in\mathbb R^n$ is an unknown disturbance, and $v_t\in\mathbb R^n$ denotes measurement noise. Moreover, $x_0 \in \mathbb R^n$ is the initial state, and $\constraintX,~\constraintU$ are the state and input constraint sets, respectively. At each time $t\in\mathbb{N}$, we only have access to the measured system state $\measx_t$. Therefore, $w_t$ captures exogenous disturbances and model mismatch, whereas $v_t$ models measurement inaccuracies. Situations where only such a noisy state $\measx_t$ is available include, e.g., (i) state estimation via an observer in case of output measurements of the form $y_t=Cx_t+Du_t$, (ii) perception-based control, where the state $x_t$ is estimated via perception maps \cite{Cothren2022,Marchi2022}, and (iii) pseudo-measurements (as frequently employed in power systems \cite{Schenato2014,Picallo2018}), all of which result in state measurement noise $v_t$. We have the following two standard assumptions on the disturbances and the system.
\begin{ass} \label{ass:noise}
	There exist $\mathcal W,\mathcal V \subseteq\mathbb R^n$ such that $v_t\in\mathcal V$ and $w_t\in \mathcal W$ hold for all $t\in\mathbb{N}_{\geq0}$. Furthermore, the sets $\mathcal V$ and $\mathcal W$ are compact, convex, and contain $\zero$ in their interior.
\end{ass}
\begin{ass} \label{ass:system}
	The pair $(A,B)$ is controllable and the sets $\constraintX$ and $\constraintU$ are compact, convex, and contain $\zero$ in their interior.
\end{ass}
The goal is to design an algorithm that achieves satisfactory performance with respect to the optimal control problem 
\begin{equation}
	\min_{\{u_t\}_{t=0}^T} \sum_{t=0}^T L_t(x_t,u_t) \quad \text{s.t. } \eqref{eq:sys_dyn}, \eqref{eq:constraints} \label{eq:OCP}
\end{equation}
for any (unknown) sequence of disturbances $\{w_t\}_{t=0}^T\in\mathcal W^{T+1}$ and despite only having access to the measured system state $\measx_t$. However, the cost functions $L_t$ are time-varying and a priori unknown, making the optimal solution inaccessible.  In particular, at each time step $t\in\mathbb{N}$, the algorithm
\begin{enumerate}
	\item obtains the noisy system state  $\measx_t$,
	\item computes a control input $u_t$ based on past measurements and cost functions, and applies it to system \eqref{eq:sys}, and
	\item receives the current cost function $L_t:\mathbb{R}^n\times\mathbb{R}^m\mapsto \mathbb{R}$.
\end{enumerate}
As detailed above, cost functions that are revealed sequentially arise frequently in various applications due to, e.g., time-varying parameters in the cost functions or tracking of an a priori unknown reference signal. Furthermore, this problem fits the OCO framework with the additional difficulty of including an underlying dynamical system. As standard in the literature on OCO-based control, we assume some regularity of the cost functions $L_t$ \cite{Agarwal2019,Li2019,Nonhoff2022a}. To this end, we define the state-input pair $z_t := (x_t,u_t)$, the constraint set $\constraintZ := \{z=(x,u): x\in\constraintX,~u\in\constraintU\}$, the closed-loop constraint set $\constraintZ_K := \{z=(x,u):x\in\constraintX,~u+Kx~\in\constraintU\}$, and the closed-loop cost functions $L_{K,t}(x,u):=L_t(x,u+Kx)$. Therein, $K\in\mathbb{R}^{m\times n}$ is an arbitrary state-feedback matrix.
\begin{ass} \label{ass:cost_function}
	For all $t\in\mathbb{N}$, the cost functions $L_t:\constraintZ \mapsto \mathbb R$ are Lipschitz continuous on $\constraintZ$ with Lipschitz constant $G$, i.e.,
	\[
	\norm{L_t(z_1) - L_t(z_2)} \leq G\norm{z_1-z_2}.
	\]
	holds for all $z_1,z_2\in\constraintZ$.
	Furthermore, the closed-loop cost functions $L_{K,t}:\constraintZ_K \mapsto \mathbb R$ are $\alpha_K$-strongly convex and have an $l_K$-Lipschitz continuous gradient on $\constraintZ_K$, i.e.,
	\begin{align*}
		L_{K,t}(z_1) &{-} L_{K,t}(z_2)  {\geq} \nabla L_{K,t}(z_2)^\top (z_1{-}z_2) {+} 	\frac{\alpha_K}{2}\norm{z_1{-}z_2}^2, \\
		&\norm{\nabla L_{K,t}(z_1) - \nabla L_{K,t}(z_2)} \leq l_K \norm{z_1-z_2}.	
	\end{align*}
	 hold for some $\alpha_K>0$, $l_K>0$, and all $z_1,z_2\in\constraintZ_K$.	
\end{ass}
Note that the constraint set $\constraintZ$ is compact by Assumption~\ref{ass:system}. Hence, it is easy to verify that Assumption~\ref{ass:cost_function} is satisfied, e.g., if the cost functions $L_t$ are $\alpha$-strongly convex on $\mathcal{Z}$, and both the cost functions $L_t$ as well as their gradients are locally Lipschitz continuous.

\section{Algorithm} \label{sec:algo}

In this section, we introduce the proposed algorithm for online convex optimization of constrained and uncertain LTI systems~\eqref{eq:sys}. Since the cost functions $L_t$ are revealed sequentially as described above, the optimal solution to~\eqref{eq:OCP} is intractable online. Instead, we propose an algorithm that aims to approximate the optimal performance by tracking the a priori unknown and time-varying optimal steady states of system \eqref{eq:sys} using the OCO framework. This strategy is in line with other works on OCO-based control and feedback optimization, compare, e.g., \cite{Lawrence2021,Colombino2020,Nonhoff2025}, and is expected to yield satisfactory performance when steady-state operation is optimal \cite{Mueller2015,Angeli2012}, and when the time-varying cost functions are changing slowly. Additionally, we develop a constraint tightening approach inspired by the robust MPC approach in \cite{Chisci2001} to cope with the constraints and disturbances acting on system~\eqref{eq:sys}. In the following, we first introduce relevant concepts for the proposed constraint tightening approach. Then, we introduce Algorithm~\ref{alg} together with relevant notation. Finally, we provide a detailed interpretation of each step and discuss implementational aspects.

Similar to \cite{Chisci2001}, we use a robust positively invariant (RPI) set for the constraint tightening.
\begin{defn}
	The set $\RPI$ is an RPI set for a system \linebreak $x_{t+1} = Ax_t+\omega$, $\omega\in\mathcal W$, if $A\RPI \oplus \mathcal W \subseteq \RPI$.
\end{defn}
Since existence of an RPI set requires system~\eqref{eq:sys} to be stable \cite{Rakovic2006}, we use a stabilizing feedback $K\in\mathbb R^{n\times m}$, which ensures that $A_K:=A+BK$ is Schur stable, i.e., the spectral radius of $A_K$ satisfies $\rho(A_K)<1$. Such a stabilizing feedback always exists by Assumption~\ref{ass:system}. Furthermore, we define the sets
\[
	\setV:= \mathcal V\oplus(-A\mathcal V) \quad \text{and} \quad \setW := \setV\oplus\mathcal W.
\] 
We use the set $\setW$ (instead of $\mathcal V \oplus \mathcal W$) in the proposed constraint tightening, because Algorithm~1 only has access to the measured system state $\measx_t$, which evolves according to
\begin{equation} \label{eq:meas_state_dynamics}
	\measx_{t+1} = x_{t+1}+v_{t+1} \refeq{\eqref{eq:sys_dyn}} A\measx_t + Bu_t + \noise_{t+1},
\end{equation}
where we define $\noise_t := w_{t-1}+v_t-Av_{t-1}\in\setW$ for all \linebreak $t\in\mathbb{N}_{\geq1}$. Next, we let $\RPI\subseteq\constraintX$ be an RPI set of the system $\chi_{t+1} = A_K\chi_t + \omega_t$, where $\omega_t \in \setW$ for all $t\in\mathbb{N}$. The corresponding minimal RPI set is given by $\RPI^*:=\sum_{i=0}^\infty A_K^i \setW$ \cite{Kolmanovsky95}. Finally, we denote by $\RPI_\mu^*:=\sum_{i=0}^\infty A_K^{\mu+i} \setW$ the minimal RPI set for the system $\chi_{t+1} = A_K\chi_t + \omega_t$, where $\omega_t \in A_K^\mu\setW$ for all $t\in\mathbb{N}$. Note that the above definitions imply
\begin{equation} \label{eq:RPI_subset}
	\RPI_\mu^* \oplus \sum_{i=0}^{\mu-1} A_K^i \setW=\RPI^*\subseteq\RPI.
\end{equation}
We note that the minimal RPI set $\RPI^*_\mu$ is convex by Assumption~\ref{ass:noise} and because $A_K$ is Schur stable \cite{Rakovic2006}. 

Furthermore, let $G_K := (I - A_K)^{-1}B$ be the map\footnote{Since $A_K$ is Schur stable, the inverse exists and the map is unique.} from an input $u^s\in\mathbb R^m$ to the corresponding steady state $x^s\in\mathbb R^n$. Then, we define the tightened set of feasible steady states by 
\[
	\feasibleSS := \{(x,u): x = G_Ku, x\in\constraintX\ominus\RPI, u+Kx\in\constraintU\ominus K\RPI\}
\]
This set of steady states is designed to ensure robust feasibility, i.e., for any $\chi_0\in\{x^s\}\oplus\RPI$, where $(x^s,u^s)\in\feasibleSS$, we have $\chi_{t+1} = A_K\chi_t+Bu^s+w_t\in\{x^s\}\oplus\RPI\subseteq\mathcal{X}$ for all $t\in\mathbb{N}$.

\begin{algorithm}[t] %
	\caption{Robust OCO for control}%
	\begin{algorithmic}
		\STATE \textbf{Initialization:} Step size $\gamma\in\left(0,\frac{2}{\alpha_K+l_K}\right]$, stabilizing feedback matrix $K\in\mathbb R^{m\times n}$, parameter $c_g\geq\norm{\controlMatrix^\top\left(\controlMatrix\controlMatrix^\top\right)^{-1}}$, prediction horizon \mbox{$\mu\geq\mu^*$}, and an initialization $\ssinput_0=\estu_0$, $\estz_0=(\estx_0,\estu_0)\in\stronglyfeasibleSS$ such that $\predu_0=\one_\mu\otimes\estu_0+\controlMatrix^\top\left(\controlMatrix\controlMatrix^\top\right)^{-1}A_K^\mu(\estx_0-\measx_0)\in\feasiblesequence(\measx_0)$.
		\STATE \textbf{At $t=0$:} Apply $u_0=T_1\predu_0+K\measx_0$
		\STATE \textbf{At each time $t\in\mathbb{N}_{\geq1}$:}
		\STATE [S1] Prediction:
		\begin{equation}
			\predx_t = A_K^\mu \measx_t + \controlMatrix \begin{bmatrix} \sigma \predu_{t-1} \\  \ssinput_{t-1} \end{bmatrix} \label{algo:prediction}\tag{A1}
		\end{equation}
		\vspace{-\baselineskip}
		\STATE [S2] Online Gradient Descent:
		\begin{align}
			&\predz_t = \begin{bmatrix} \predx_t \\ \ssinput_{t-1} \end{bmatrix} \label{algo:def_predz} \tag{A2} \\
			&\estz_t = \begin{bmatrix} \estx_t \\ \estu_t \end{bmatrix} = \Pi_{\stronglyfeasibleSS} \left( \predz_t - \gamma \overbar{K}^\top \nabla L_{t-1} \left( \predx_t, \ssinput_{t-1}+K\predx_t \right) \right) \label{algo:OGD}\tag{A3}
		\end{align}
		\vspace{-\baselineskip}
		\STATE [S3] Additional Input Sequence:
			\begin{align}
				&\text{Find $g_t\in\mathbb R^{\mu m}$ s.t.} \nonumber \\
				&\qquad \controlMatrix g_t = \estx_t - \predx_t,\quad \norm{g_t}\leq c_g \norm{\estx_t-\predx_t}
				\label{algo:additional_input}\tag{A4}
		\end{align}
		\vspace{-\baselineskip}
		\STATE [S4] Predicted Input Sequence:
			\begin{align}
				&\beta_t = \max_{\beta\in[0,1]} \beta~\text{s.t.}~\begin{bmatrix} \sigma\predu_{t-1} \\ \ssinput_{t-1} \end{bmatrix} + \beta g_t\in\feasiblesequence(\measx_t) \label{algo:beta} \tag{A5} \\
				&\predu_t = \begin{bmatrix} \sigma \predu_{t-1} \\ \ssinput_{t-1} \end{bmatrix} + \beta_t g_t \label{algo:predicted_input}\tag{A6} \\
				&\ssinput_t = (1-\beta_t) \ssinput_{t-1} + \beta_t \estu_t \label{algo:us_t}\tag{A7}			
		\end{align}
		\vspace{-\baselineskip}
		\STATE [S5] Control Input:
		\begin{equation}			
			u_t = T_1\predu_t + K\measx_t \label{algo:control_input}\tag{A8}
		\end{equation}
		\vspace{-\baselineskip}
		\STATE [S6] Apply $u_t$ to system \eqref{eq:sys}, receive the cost function $L_t(u,x)$, and move to [S1] at time $t+1$
	\end{algorithmic}
	\label{alg}
\end{algorithm}

Next, we introduce Algorithm~\ref{alg}, which is illustrated graphically in Figure~\ref{fig:schematic}. Algorithm~\ref{alg} aims to estimate and track the time-varying optimal steady states of system~\eqref{eq:sys} given by
\begin{align}
	(\theta_t,\eta_t) := \arg\min_{(x,u)\in\stronglyfeasibleSS} L_{K,t}(x,u), \label{eq:optimal_steady_state}
\end{align}
which are unique due to strong convexity of the cost functions $L_{K,t}$. As standard in constrained tracking control (e.g., \cite[Assumption 1]{Koehler2020}), we require that the references, i.e., the optimal steady states, strictly satisfy the constraints $(\theta_t,\eta_t)\in\stronglyfeasibleSS$, where $\stronglyfeasibleSS\subseteq\text{rel\,int }\feasibleSS$ is compact, convex, and contains $\zero$ in its relative interior. For compactness, we abbreviate $\zeta_t := \begin{bmatrix} \theta_t^\top & \eta_t^\top \end{bmatrix}^\top$.

In order to estimate and track the time-varying optimal steady states $(\theta_t,\eta_t)$, Algorithm~\ref{alg} repeats the following steps at each time step $t\in\mathbb{N}$:
	
First, the noisy system state $\measx_t$ is measured. Then, Algorithm~\ref{alg} computes a $\mu$-step ahead prediction of the system state $\predx_t$ based on the measurement $\measx_t$ and a $\mu$-step sequence of predicted control inputs computed at the previous time step (trajectory \protect\circled{1} in Figure~\ref{fig:schematic}) in [S1]. Therein, \linebreak $\controlMatrix := \begin{bmatrix} A_K^{\mu-1}B & A_K^{\mu-2}B & \dots & B \end{bmatrix}$ is the controllability matrix\footnote{We change the order of entries in the matrix $\controlMatrix$ compared to the standard definition, so that the first entry $u_1\in\mathbb R^m$ of the input sequence $u=\begin{bmatrix} u_1^\top & \dots & u_\mu^\top \end{bmatrix}$ is the first (in time) input applied in the $\mu$-step ahead prediction $x_{t+\mu} = A_K^\mu x_t + \controlMatrix u$.} of the stabilized system, $\mu\geq\mu^*$ is the prediction horizon of Algorithm~\ref{alg}, and $\mu^*\in\mathbb{N}$ is the controllability index, i.e., the smallest integer such that $\text{rank }\controlMatrix=n$. Next, we apply projected online gradient descent (OGD) \cite{Zinkevich03} to get an estimate $\estz_t$ of the optimal steady state in step [S2], where $\overbar{K}:=\begin{bmatrix} I_n & \zero_{n,m} \\ K & I_m \end{bmatrix}$. In~[S3], an additional input sequence $g_t$ is calculated that steers the system to the estimated optimal steady state $\estx_t$ in $\mu$ steps while neglecting the constraints (trajectory \protect\circled{2} in Figure~\ref{fig:schematic}). Note that such a sequence exists by Assumption~\ref{ass:system} and because $\mu\geq\mu^*$. Then, the predicted input sequence $\predu_t$ is computed by scaling the additional input sequence $\beta_tg_t$. The scaling $\beta_t$ in [S4] is crucial in order to enforce closed-loop constraint satisfaction. To this end, the constraint set $\feasiblesequence(x)$ is defined by
\begin{align}
\begin{split} \label{eq:tightened_constraint_set}
	&\feasiblesequence(x) := \Big\{u\in\mathbb R^{\mu m}: ~\forall \tau\in\mathbb{N}_{[0,\mu-1]}:~x_0 = x, \\
	&x_{\tau+1} {=} A_K^{\tau+1}x_0 {+} \sum_{i=0}^\tau A_K^i B T_{\tau-i+1} u \in \constraintX \ominus \sum_{j=0}^{\tau} A_K^j\setW,\\
	&T_{\tau+1} u + Kx_\tau \in \constraintU {\ominus} K \sum_{j=0}^{\tau-1} A_K^{j}\setW \Big\}.
\end{split}
\end{align}
In particular, the tightened constraints $\feasiblesequence(x)$ are designed such that the original constraints~\eqref{eq:constraints} are tightened more the farther Algorithm~\ref{alg} predicts into the future in order to cope with the growing uncertainty (in time) due to the disturbances acting on the system (cf.~\cite{Chisci2001} for a similar constraint tightening approach in the context of model predictive control). However, due to the scaling $\beta_t$, the predicted input sequence $\predu_t$ does not steer the system to the estimated optimal steady state $\estx_t$ anymore. Instead, $\predu_t$ steers the system to a convex combination of the prediction $\predx_t$ and the estimate $\estx_t$, i.e., to the vicinity of a different steady state $\ssstate_t$ (trajectory \protect\circled{3} in Figure~\ref{fig:schematic}). We keep track of the corresponding steady-state input $\ssinput_t$ in [S4] and define $\ssstate_t:=G_K\ssinput_t$ and $\ssz_t = \begin{bmatrix} \left(\ssstate_t\right)^\top & \left( \ssinput_t \right)^\top \end{bmatrix}^\top$. Finally, the first part of the predicted input sequence and the stabilizing feedback $K\measx_t$ are applied to system~\eqref{eq:sys} in steps [S5] and [S6]. After applying $u_t$, we receive the cost function $L_t(x,u)$, and repeat the procedure at the next time step $t+1$. 

\begin{figure}
	\centering
	\begin{tikzpicture}[scale=1.0]
	\definecolor{mygreen}{rgb}{0.2,0.7,0.2}
	\clip (0.19,-1.01) rectangle (8.51,5.51);
	
	\draw[thick] (0.2,-1) rectangle (8.5,4.75);
	\node at (8.3,-0.8) {$\mathcal X$};
	\draw[thick] (0.6,-0.8) rectangle (8.1,4.55);
	\node at (6.8,-0.55) {$\mathcal X\ominus\setW$};
	\draw[thick] (1.5,0) -- (6.7,0) -- (7.7,1) -- (7.7,3.05) -- (7.2,3.75) -- (2,3.75) -- (1,2.75) -- (1,0.7) -- (1.5,0);
	\node at (6.75,0.65) {$\mathcal X\ominus\RPI$};
	\draw[red,dotted,thick] (1.25,2.5625) -- (4.5,0.125); 
	
	\node at (4.6,0.3) {$\bar{\mathcal{S}}$};
	
	\coordinate (theta) at (2,3.95);					
	\draw[dotted,rotate around={100:(theta)}] (theta) ellipse (2 and 1);
	\draw[dotted,rotate around={100:(theta)}] (theta) ellipse (4 and 2);
	\draw[dotted,rotate around={100:(theta)}] (theta) ellipse (6 and 3);
	\draw[dotted,rotate around={100:(theta)}] (theta) ellipse (8 and 4);
	\draw[dotted,rotate around={100:(theta)}] (theta) ellipse (10 and 5);
	
	\node at (1.75,2.1875) {\tiny$\times$};
	\node[inner sep = 0] at (1.6,2) {$\theta_{t-1}$};
	
	\coordinate (yt-1) at (8.3,4);
	\node[fill=white,inner sep = 1] at (8.15,4.3) {$\measx_{t}$};
	\node at (8.3,4) {\tiny$\times$};
			
	\node[inner sep = 0] (haty) at (3.9,0.3) {\tiny$\times$};
	\node[fill=white,inner sep = 0] at (3.6,0.1) {$\predx_t$};			
	\draw (yt-1) -- (7,2.25) -- (6,1.25) --node[midway,above,circle,draw, inner sep = 1,yshift=2] {\tiny{1}} (5,0.75) -- (haty.center) ;
	\node[inner sep = 0] at (3.1408,1.1444) {\tiny$\times$};
	\node[fill=white,inner sep = 0] at (3.4,1.2) {$\ssstate_t$};		
	
	\coordinate (grad) at (2.5,1.05);
	\node[inner sep = 0] (ys) at (2.776,1.418) {\tiny$\times$};
	\draw[thick,->,blue,dashed] (haty) -- (grad);
	\draw[thick, dashed,blue] (grad) -- (ys.center);
	\node[fill=white,inner sep = 0,xshift=-3,yshift=3] at (ys.west) {$\estx_t$};
	\node[fill=white,inner sep = 0] at (2.5,0.5) {$-\gamma\nabla L_{K,t-1}$};
	
	\draw (yt-1) -- (6,5.5) -- (4.25,5) --node[circle,draw, inner sep = 1,midway,above,xshift=-4] {\tiny{2}} (3,3.5) -- (ys.center);
	\draw[dashed, gray] (haty) -- (ys);
		
	\coordinate (ytmeas) at (6.29,4.55); 
	\draw[thick,mygreen] (yt-1) -- (ytmeas);
	\draw (ytmeas) -- (4.76,3.91) --node[circle,draw, inner sep = 1,pos=0.35,below,yshift=-3] {\tiny{3}} (3.58,2.7) -- (3.1,1.09);
	\draw[dashed, gray] (7,2.25) -- (6,5.5);
	\draw[dashed, gray] (6,1.25) -- (4.25,5);
	\draw[dashed, gray] (5,0.75) -- (3,3.5);
	\node[gray] at (3.6,3.3) {\footnotesize$1{-}\beta_t$};
	\node[gray] at (4.4,2) {\footnotesize$\beta_t$};
\end{tikzpicture}
\vspace{-3ex}
	\caption{Schematic illustration of Algorithm~\ref{alg}: The closed-loop cost function $L_{K,t-1}$ is visualized via its sublevel sets (dotted) together with the (tightened) constraint sets and the tightened steady-state manifold $\stronglyfeasibleSS$ (red, dotted). Note that the constraints are tightened the farther Algorithm~\ref{alg} predicts into the future. First, Algorithm~\ref{alg} predicts the system state $\mu$ times state ahead $\predx_t$ \protect\circled{1}. Then, it applies OGD evaluated at $\predx_t$ in [S2] (blue, dashed) to obtain an estimate of the optimal steady state $\estx_t$. Next, an additional input sequence $g_t$ is computed that steers the closed-loop system to the estimate $\estx_t$ \protect\circled{2}. Note that the additional input sequence $g_t$ may violate the constraints. Therefore, a scaling $\beta_t$ (gray, dashed) is applied to ensure that the trajectory emerging from application of the predicted input sequence $\predu_t$ \protect\circled{3} robustly satisfies the constraints. However, due to the scaling $\beta_t$, the predicted input sequence steers the system to a vicinity of the steady state $\ssstate_t$ instead of to the estimate $\estx_t$. Finally, the first part of the predicted input sequence is applied to system~\eqref{eq:sys}.}
	\label{fig:schematic}
\end{figure}

\noindent\textbf{Description and interpretation of Algorithm~\ref{alg}.}

[S1] First, Algorithm~\ref{alg} predicts the system state $\mu$ time steps ahead based on the measurement $\measx_t$ and the stabilized system dynamics. As discussed above, the previously predicted input sequence $\predu_{t-1}$ steers the system to a vicinity of the steady state $\ssstate_{t-1}$. Hence, we append the shifted previously predicted input sequence $\sigma\predu_{t-1}$ with the steady-state input $\ssinput_{t-1}$ to ensure that the predicted state $\predx_t$ remains in a vicinity of $\ssstate_{t-1}$, which guarantees robust constraint satisfaction in our analysis below.

[S2] Next, Algorithm~\ref{alg} computes an estimate of the optimal steady state by applying OGD based on the previous cost function accounting for the cost of the stabilizing input, i.e., based on $L_{K,t-1}$. Note that, due to the chain rule, we have $\overbar{K}^\top\nabla L_{t-1}(x,u+Kx) = \nabla L_{K,t}(x,u)$. We note that different procedures to obtain the estimate $\estz_t$ in [S2] are possible. More specifically, we only require that $\norm{\estz_t - \zeta_{t-1}} \leq \kappa \norm{\predz_t - \zeta_{t-1}}$ holds for some $\kappa\in[0,1)$ and all $t\in\mathbb{N}_{\geq1}$ in our analysis below. In particular, algorithms that achieve this condition include OGD \cite{Zinkevich03} due to Assumption~\ref{ass:cost_function} (compare~\eqref{eq:GD_contr}) and many first-order optimization methods \cite{Necoara2019}. Alternatively, computing the previously optimal steady state exactly and setting $\estz_t = \zeta_{t-1}$ satisfies the condition above with $\kappa=0$.

[S3] In order to steer the stabilized system to the estimated optimal steady state $\estx_t$, Algorithm~\ref{alg} computes an additional input sequence $g_t$ next. In particular, $g_t$ is computed such that $A_K^\mu \measx_t + \controlMatrix \begin{bmatrix} \left(\sigma\predu_{t-1}\right)^\top & \left(\ssinput_{t-1}\right)^\top\end{bmatrix}^\top + S_c g_t = \predx_t + S_c g_t = \estx_t$. In particular, we require that the prediction horizon satisfies $\mu\geq\mu^*$, and we assume that the system dynamics~\eqref{eq:sys_dyn} are controllable in Assumption~\ref{ass:system} to ensure that the estimated optimal steady state is always reachable in $\mu$ steps when neglecting the constraints. The additional inequality constraint in~\eqref{algo:additional_input} is needed to avoid solutions of~\eqref{algo:additional_input} that yield poor performance: Consider the case that both Algorithm~\ref{alg} and system~\eqref{eq:sys} have converged to the optimal steady state, i.e., $x_t = \predx_t = \estx_t = \theta_{t-1}$ and $T_\tau \begin{bmatrix} \left(\sigma\predu_{t-1}\right)^\top & \left(\ssinput_{t-1}\right)^\top\end{bmatrix}^\top=\eta_{t-1}$ for any $\tau\in\mathbb{N}_{[1,\mu]}$. Without the additional constraint, it would be possible to choose an input sequence $g_t$ such that $A_Kx_t+BT_1\left(\sigma\predu_{t-1}\right)+BT_1g_t\neq\theta_{t-1}$, but $S_c g_t =  \estx_t-\predx_t$, i.e., this input sequence would be such that the system initially moves away from the optimal steady state, but then converges again within $\mu$ time steps. Such an input sequence would yield poor performance, and is therefore undesirable in practice. In contrast, the additional constraint enforces $g_t = \zero$ in this case. Finally, note that different solutions to the feasibility problem in~\eqref{algo:additional_input} are possible within Algorithm~\ref{alg}. In particular,~\eqref{algo:additional_input} admits an explicit solution\footnote{The inverse exists because $\controlMatrix$ has full row rank due to Assumption~\ref{ass:system} and $\mu\geq\mu^*$.} $g_t = \controlMatrix^\top \left( \controlMatrix\controlMatrix^\top \right)^{-1} \left( \estx_t - \predx_t \right)$. Alternatively, it is possible to compute a (sub)optimal solution with respect to an appropriately chosen cost criterion in~\eqref{algo:additional_input} to improve transient performance and allow implementation of soft constraints (cf. Section~\ref{sec:sim}), albeit at the cost of increased computational complexity.

[S4] Next, Algorithm~\ref{alg} computes a scaling $\beta_t\in[0,1]$ and the predicted input sequence $\predu_t$. The scaling $\beta_t$ is crucial to guarantee robust constraint satisfaction for the closed loop in our analysis below. However, as discussed above, the scaling results in the predicted input sequence $\predu_t$ steering the system to a vicinity of the steady state $\ssstate_t$. Hence, we additionally compute the corresponding steady-state input $\ssinput_t$, which we require for the prediction step [S1] at the next time step.

[S5] Finally, Algorithm~\ref{alg} combines the first part of the predicted input sequence $T_1\predu_t$ with the stabilizing feedback $K\measx_t$ to compute the control input $u_t$.

\noindent\textbf{Implementational aspects of Algorithm~\ref{alg}.}

In Algorithm~\ref{alg}, design variables for tuning the algorithm are the stabilizing feedback $K$, the prediction horizon $\mu$, the parameter $c_g$, the step size for OGD $\gamma$, and the tightened steady-state manifold $\stronglyfeasibleSS$. We found in simulations that the effect of the stabilizing feedback on the closed loop performance is small. Therefore, $K$ should be chosen to minimize the constraint tightening \eqref{eq:tightened_constraint_set}. The prediction horizon $\mu$ needs to be larger than or equal to the controllability index $\mu^*$. Furthermore, a smaller prediction horizon forces Algorithm~\ref{alg} to satisfy the equality constraint in~\eqref{algo:additional_input} in shorter time. Therefore, a smaller prediction horizon generally makes the closed loop more aggressive, at the cost of larger control inputs. The parameter $c_g$ can be neglected if~\eqref{algo:additional_input} is solved explicitly. If~\eqref{algo:additional_input} is solved by optimization, then $c_g$ should be chosen large enough. In particular, $c_g \geq \norm{\controlMatrix^\top\left(\controlMatrix\controlMatrix^\top\right)^{-1}}$ is required in order to ensure that a feasible candidate solution to~\eqref{algo:additional_input} exists at all times $t\in\mathbb{N}$ (given by the explicit solution discussed above). 
The step size parameter $\gamma$ needs to be chosen from the interval $(0,\frac{2}{\alpha_K+l_K}]$ and, similar to the prediction horizon $\mu$, can be tuned to achieve a satisfactory tradeoff between convergence speed and size of the control inputs. The tightened steady-state manifold $\stronglyfeasibleSS$ should typically be chosen as large as possible, e.g., $\stronglyfeasibleSS=(1-\delta_s)\feasibleSS$, where $\delta_s\in(0,1)$ is some arbitrary small constant.

The main computational burden of Algorithm~\ref{alg} are the projection onto $\stronglyfeasibleSS$ in~\eqref{algo:OGD} and the feasibility problem in~\eqref{algo:additional_input}. Typically, the steady-state manifold $\feasibleSS$ is low-dimensional. Additionally, the set $\stronglyfeasibleSS$ can be chosen to further simplify the set $\feasibleSS$, thereby enabling an efficient implementation of the projection at the cost of closed-loop performance. The feasibility problem in~\eqref{algo:additional_input} can either be solved explicitly or be formulated as an optimal control problem to improve transient performance, at the cost of increased computational complexity. The scalar optimization problem~\eqref{algo:beta} can be solved efficiently via bisection. The following analysis of Algorithm~\ref{alg} in Section~\ref{sec:results} is independent of how a solution to~\eqref{algo:additional_input} is chosen. 

\section{Theoretical Results} \label{sec:results}

In this section, we derive theoretical guarantees for Algorithm~\ref{alg}. All proofs are deferred to the appendix. We show that Algorithm~\ref{alg} is well-defined, i.e., that the feasibility problem~\eqref{algo:additional_input} and the optimization problem~\eqref{algo:beta} always admit feasible solutions, that it guarantees robust constraint satisfaction for system~\eqref{eq:sys}, and achieves bounded dynamic regret.

First, it is of paramount importance to ensure that the control input $u_t$ provided by Algorithm~\ref{alg} exists and is well-defined for all $t\in\mathbb{N}$, i.e., that feasible solutions to~\eqref{algo:additional_input} and~\eqref{algo:beta} exist at all times. To this end, we assume that the initialization of Algorithm~\ref{alg} satisfies the tightened constraints as follows.
\begin{ass} \label{ass:init}
	The initialization of Algorithm~\ref{alg} satisfies $\predu_0 \in \feasiblesequence(\measx_0)$ and $\estz_0\in\stronglyfeasibleSS$. Furthermore, $x_0\in\constraintX$.
\end{ass}
Using Assumption~\ref{ass:init}, we can show existence of the desired solutions and robust constraint satisfaction in the next lemma.
\begin{lem} \label{lem:feasibility}
	Suppose Assumptions~\ref{ass:noise}, \ref{ass:system}, and \ref{ass:init} are satisfied. Let $\mu\geq\mu^*$ and $c_g\geq \norm{\controlMatrix^\top\left(\controlMatrix\controlMatrix^\top\right)^{-1}}$. It holds that
	\begin{itemize}
		\item[(i)] there exist feasible solutions to~\eqref{algo:additional_input}-\eqref{algo:beta} for all $t\in\mathbb{N}_{\geq1}$,
		\item[(ii)] $\begin{bmatrix} (\sigma\predu_{t-1})^\top & (\ssinput_{t-1})^\top \end{bmatrix}^\top \in\feasiblesequence(\measx_t)$ for all $t\in\mathbb{N}_{\geq1}$, and $\predu_t\in\feasiblesequence(\measx_t)$ for all $t\in\mathbb{N}$,
		\item[(iii)] $\ssz_t=(\ssstate_t,\ssinput_t) \in \stronglyfeasibleSS$ for all $t\in\mathbb{N}$,
		\item[(iv)] $\predx_{t+1}\in\{\ssstate_t\}\oplus\RPI_\mu^*$ for all $t\in\mathbb{N}$, and
		\item[(v)] $x_t\in\constraintX$ and $u_t\in\constraintU$ for all $t\in\mathbb{N}$.
	\end{itemize}
\end{lem}
The proof is given in Appendix \ref{sec:proofA}. 

Next, in order to prove bounded dynamic regret, we need to ensure that Algorithm~\ref{alg} responds to changes in the cost function and achieves a certain rate of convergence. 
\begin{lem} \label{lem:mincontr}
	Suppose Assumptions~\ref{ass:noise}, \ref{ass:system}, and \ref{ass:init} are satisfied. Let $\mu\geq\mu^*$ and $c_g\geq\norm{\controlMatrix^\top\left(\controlMatrix\controlMatrix^\top\right)^{-1}}$. There exists $b \in [0,1)$ such that $\prod_{k=0}^\mu (1-\beta_{t+k}) \leq b$ holds for all $t\in\mathbb{N}_{\geq1}$.
\end{lem}
The proof of Lemma~\ref{lem:mincontr} is detailed in Appendix~\ref{sec:proofB}. Lemma~\ref{lem:mincontr} guarantees that, over a horizon of $\mu$ time steps, the scaling $\beta_t$ cannot be equal to $0$ at all times. Since the predicted input sequence $\predu_t$ is adapted to track the optimal steady state whenever $\beta_t\neq0$ (compare \eqref{algo:predicted_input}), Lemma~\ref{lem:mincontr} can be viewed as a result on the average convergence of the closed loop, thereby enabling us to study the closed-loop performance of Algorithm~\ref{alg} next. To this end, we employ dynamic regret \eqref{eq:def_regret_intro} as a measure of the proposed algorithm's performance. In this work, we employ the optimal steady states $(\theta_t,\eta_t)\in \mathbb R^{n+m}$ \eqref{eq:optimal_steady_state} as a performance benchmark. Thus, the dynamic regret of Algorithm~\ref{alg} is given by
\begin{equation}
	\regret := \sum_{t=0}^T L_t(x_t,u_t) - L_t(\theta_t,\eta_t+K\theta_t). \label{eq:def_regret}
\end{equation}
This choice of benchmark is motivated by the fact that Algorithm~\ref{alg} aims to approximate the optimal performance by tracking the optimal steady states of system~\eqref{eq:sys}. Furthermore, the optimal steady states are also commonly analyzed as a performance benchmark in, e.g., economic MPC \cite{Faulwasser18,Angeli2012} and previous works on OCO-based control \cite{Nonhoff2022a,Nonhoff2025}. As an alternative benchmark, the optimal solution to~\eqref{eq:OCP} in hindsight has been considered in the literature, compare, e.g., \cite{Li2019}. Obtaining an upper bound on the dynamic regret of Algorithm~\ref{alg} with respect to this benchmark is an interesting direction for future research, but beyond the scope of this work.

Finally, the main result of this paper provides an upper bound for the dynamic regret~\eqref{eq:def_regret} of Algorithm~\ref{alg}.
\begin{thm} \label{thm}
	Suppose Assumptions \ref{ass:noise}--\ref{ass:init} are satisfied. Let $\gamma\in\left(0,\frac{2}{\alpha_K+l_K}\right]$, $\mu\geq\mu^*$, and $c_g\geq\norm{\controlMatrix^\top\left(\controlMatrix\controlMatrix^\top\right)^{-1}}$. Then, there exists a feasible solution for~\eqref{algo:additional_input}-\eqref{algo:beta} for all $t\in\mathbb{N}_{\geq1}$, and there exist constants $c_0,c_\zeta,c_w,c_v>0$ such that
	\[
	\regret \leq c_0 + c_\zeta \sum_{t=1}^T \norm{\zeta_t - \zeta_{t-1}} + c_w \sum_{t=0}^{T-1} \norm{w_t} + c_v \sum_{t=0}^T \norm{v_t}
	\]
	holds for all $T\in\mathbb{N}_{\geq1}$, and any sequence of cost functions $L_t$ and disturbances $\{w_t\}_{t=0}^T$ and $\{v_t\}_{t=0}^T$. Moreover, the constraints are satisfied, i.e., $x_t\in\mathcal X$ and $u_t\in\mathcal U$ hold for all $t\in\mathbb{N}$.
\end{thm}
Next, we provide a brief sketch of the proof of Theorem~\ref{thm}, which is detailed in Appendix~\ref{sec:proofC}.

\textit{Sketch of proof:} Constraint satisfaction and existence of a feasible solution to~\eqref{algo:additional_input} and~\eqref{algo:beta} follow from Lemma~\ref{lem:feasibility}. Then, Lipschitz continuity in Assumption~\ref{ass:cost_function} yields
\begin{equation*}
	\begin{split}
		\regret&\leq \tilde{C}_0 + c_{GK} \underbrace{\sum_{t=1}^{T-\mu} \norm{\begin{bmatrix} \measx_{t+\mu} \\ T_1 \predu_{t+\mu} \end{bmatrix} - \predz_t}}_\text{Part III} + G\sum_{t=0}^T \norm{v_t} \\ &\quad +c_{GK} \underbrace{\sum_{t=1}^{T-\mu}\norm{\predz_t - \zeta_t}}_\text{Part II}  + c_{GK} \underbrace{\sum_{t=1}^{T-\mu} \norm{\zeta_{t+\mu} - \zeta_t}}_\text{Part I},
	\end{split}
\end{equation*}
where $\tilde{C}_0 := G\sum_{t=0}^\mu \norm{\begin{bmatrix} \measx_t-\theta_t \\ K\measx_t + T_1\predu_t - (K\theta_t+\eta_t) \end{bmatrix} }$ captures the initialization error, and $c_{GK} := G(\norm{K}+1)$. We bound these three sums separately.

\noindent\underline{\textit{Part I:}} The first part describes the regret of tracking the optimal steady states $\zeta_t$ with a $\mu$-step delay, which arises due to the $\mu$-step ahead predictions in Algorithm~\ref{alg}. It is straightforward to show that $\text{Part I}\leq\mu\sum_{t=1}^T \norm{\zeta_t-\zeta_{t-1}}$.

\noindent\underline{\textit{Part II:}} This sum describes the prediction error of Algorithm~\ref{alg}, compare~\eqref{algo:prediction} and~\eqref{algo:def_predz}. In order to derive an upper bound for this part, we first show that the predictions $\predz_t$ evolve in a tube around the steady states $\ssz_{t-1}$ given by the RPI set $\RPI$. Then, combining Lemma~\ref{lem:mincontr} and contraction properties of gradient descent (compare~\eqref{eq:GD_contr}), we obtain $\text{Part II}\leq c_0^\mathrm{II}+c_\zeta^\mathrm{II}\sum_{t=1}^{T-\mu}\norm{\zeta_t-\zeta_{t-1}}+c_w^\mathrm{II}\sum_{t=1}^{T-\mu}\norm{\noise_t}$, where $c_0^\mathrm{II},c_\zeta^\mathrm{II},c_w^\mathrm{II}>0$.

\noindent\underline{\textit{Part III:}} Finally, Part III describes the mismatch between the predictions at time $t$ and the measured system state and input at time $t+\mu$. An upper bound for this part can be derived by exploiting stability of $A_K$ and the inequality constraint in~\eqref{algo:additional_input}. Then, we get $\text{Part III}\leq c_0^\mathrm{III}+c_\zeta^\mathrm{III}\sum_{t=1}^{T-\mu}\norm{\zeta_t-\zeta_{t-1}}+c_w^\mathrm{III}\sum_{t=1}^{T-\mu}\norm{\noise_t}$, where $c_0^\mathrm{III},c_\zeta^\mathrm{III},c_w^\mathrm{III}>0$.

\noindent The desired result follows by combining these bounds.\hfill$\blacksquare$

The upper bound in Theorem~\ref{thm} depends linearly on $\sum_{t=1}^T \norm{\zeta_t - \zeta_{t-1}}$, commonly termed path length in the literature \cite{Li2021a,Karapetyan2023}, which can be interpreted as a measure for the variation of the cost functions. By adapting~\cite[Theorem~3]{Li2019}, it can be shown for the nominal setting (i.e., without disturbances) that an upper bound that depends linearly on the path length is optimal, i.e., the best achievable bound. Additionally, such a bound implies asymptotic stability of the optimal steady state (if the cost function is constant) in a similar nominal setting \cite{Nonhoff2023a}. Furthermore, the upper bound depends linearly on the magnitude of the disturbances $w_t$ and measurement noise $v_t$. Such a dependence has to be expected, because the optimal steady states, which serve as a benchmark in our definition of dynamic regret, are not affected by the disturbances. In contrast, these disturbances have the capability to drive the closed loop away from the optimal steady state, even if the cost function remains constant.

\begin{rem}
	As discussed above, our setting and hence also the regret bound from Theorem~\ref{thm} is applicable to different application scenarios considered in the literature. For example, in perception-based control \cite{Dean2021,Cothren2022,Marchi2022}, $v_t$ is the error from the perception maps, which can be further bounded if, e.g., a residual neural network is used for perception \cite{Marchi2022}. More specifically, our results are applicable to the setting considered in \cite{Dean2021}, where a tracking control problem for an LTI system with complex, nonlinear measurements is studied (note that, as a special case, our results also hold for $w_t\equiv 0$). Furthermore, or results are applicable in case that only partial feedback is available instead of full state measurements. Then, an observer can be implemented such that the measurement noise $v_t$ captures the observer error. Moreover, they are applicable to the setting in \cite{Colombino2020}, which considers optimal steady-state tracking of an LTI system. In particular, the algorithm in \cite{Colombino2020} only ensures constraint satisfaction at the optimal steady state, whereas our proposed algorithm also ensures constraint satisfaction during the transient.
\end{rem}
\section{Numerical Case Study} \label{sec:sim}

In this section, we illustrate applicability and validate our theoretical results on a numerical simulation\footnote{The code for the simulations can be found online at https://doi.org/10.25835/og5nute0.}. We apply two variants of Algorithm~\ref{alg} to a tracking control problem for an autonomous vehicle. First, we solve an optimal control problem in \eqref{algo:additional_input} with an appropriately defined cost function in order to achieve satisfactory transient performance. Second, we apply the explicit solution for~\eqref{algo:additional_input} discussed above and compare the results. The vehicle is modeled using its nonlinear kinematics
\[
	\dot x_{t} = \begin{bmatrix} \varDelta_t\cos(\delta_t) \\ \varDelta_t\sin(\delta_t) \\ a_t \end{bmatrix},
\]
where the system states are $x_t = \begin{bmatrix} p_{x,t} & p_{y,t} & \varDelta_t \end{bmatrix}^\top \in \mathbb R^3$, $p_{x,t}$ and $p_{y,t}$ are the longitudinal and lateral position at time $t$, and $\varDelta_t$ is the car's velocity. The control inputs are $u_t = \begin{bmatrix} \delta_t & a_t \end{bmatrix}^\top \in \mathbb R^2$, where $\delta_t$ is the steering angle, and $a_t$ the acceleration. First, we discretize the model using a sample time $\tau = 0.1$\,s and linearize it around $\overbar{\varDelta} = 100$\,km/h. The resulting errors are taken into account by defining $\mathcal W:= \{w\in\mathbb R^3: \norm{w}_\infty\leq0.2\}$. We constrain the lateral position $p_y$ such that the car stays on a two-lane road $p_{y,t}\in[-1.5\,\text{m},4.5\,\text{m}]$, i.e., the middle of the right lane is at $p_y=0$\,m and the middle of the left lane at $p_y=3$\,m. Moreover, the car's velocity is constrained to $\varDelta_t\in[0\,\text{km/h},130\,\text{km/h}]$, and the control inputs need to satisfy $\delta_t \in [-20^\circ,20^\circ]$ and $a_t\in[-4\,\text{m/s}^2,4\,\text{m/s}^2]$. We assume that an online planner is available, which decides on a desired behavior online and provides a corresponding cost function, thereby making the cost functions a priori unknown and time-varying. 

Next, we implement Algorithm~\ref{alg} to control the autonomous car. If we applied Algorithm~\ref{alg} to the full system, the steady state manifold $\mathcal S$ in \eqref{algo:OGD} would only include states $x_t$ such that $\varDelta_t=0$, which leads to undesirable behavior (i.e., the car almost stopping). Therefore, we only apply Algorithm~1 to the lateral position $p_{y,t}$ and velocity $\varDelta_t$. This is possible, because the longitudinal position $p_{x,t}$ does not affect the other states and there are no constraints acting on it. Thus, we compute a stabilizing feedback $K$ for these linearized reduced dynamics. For computation of the RPI set $\RPI$, we use the method described in \cite{Rakovic2005} and the multi-parametric toolbox 3 \cite{MPT3}. In Algorithm~\ref{alg}, we set $\bar{\mathcal S} = 0.99\mathcal S$, $\gamma=0.7$, $c_g = 1000$, and $\mu=10$, i.e., the prediction horizon is set to $1$\,s. First, we solve \eqref{algo:additional_input} by optimizing a cost function $L^g_t$ defined below subject to the constraints in \eqref{algo:additional_input}. The optimization problem is solved using the Casadi toolbox \cite{casadi}, whereas we solve~\eqref{algo:beta} via bisection. Algorithm~\ref{alg} and the autonomous car are initialized traveling on the right lane with a constant speed $\varDelta_0=120$\,km/h. Furthermore, we place another slower vehicle on the right lane, which is $150$\,m ahead of the controlled car and traveling with a constant speed $\varDelta^c = 70$\,km/h. Finally, we assume that the autonomous car is equipped with sensors that can measure the controlled car's velocity, lateral position, and distance $d_t$ to the vehicle ahead. For each sensor, we add measurement noise sampled randomly uniformly from the intervals $[-0.1\,\text{m},0.1\,\text{m}]$ and $[-0.1\,\text{km/h},0.1\,\text{km/h}]$.

\begin{figure}
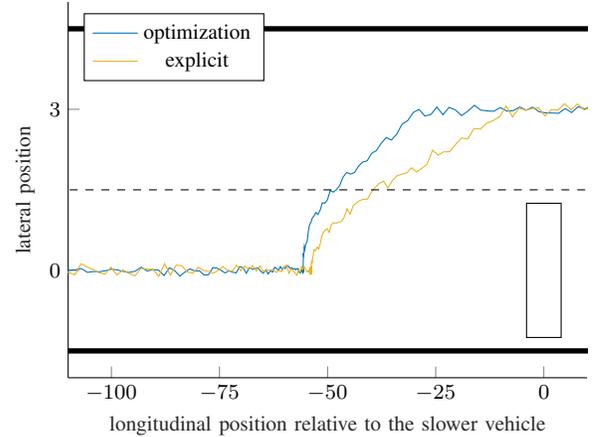

	\centering
	\footnotesize
	\setlength{\hohe}{5cm}
	\setlength{\breite}{.4\textwidth}
	\include{figures/scenario}
	\vspace{0cm}
	\caption{Schematic illustration of the scenario considered in the simulation. The border of the road is indicated by the thick black lines, the two lanes are illustrated by the dashed line. The position of the controlled car relative to the slower vehicle is shown in blue (for the optimization based solution of \eqref{algo:additional_input}) and yellow (for the explicit solution of \eqref{algo:additional_input}). The slower vehicle is indicated by the black rectangle.}
	\label{fig:scenario}
\end{figure}

Our simulation can then be separated into three phases. A graphical illustration of the scenario is given in Figure~\ref{fig:scenario}.
\begin{enumerate}
	\item Initially, the controlled car is unaware of the slower moving vehicle in front. The planner therefore decides on a constant cost function $L_{p1}(x,u;\theta_{p1}^y,\theta_{p1}^\varDelta)=\frac{1}{2}\norm{p_y-\theta_{p1}^y}^2 + \frac{1}{2} \norm{\varDelta-\theta_{p1}^\varDelta}^2 + 50 \norm{u}^2$, where $\theta_{p1}^y = 0$\,m and $\theta_{p1}^\varDelta = 120$\,km/h. Thus, the car is operated optimally when driving on the right lane with a constant speed equal to $\theta_{p1}^v=120$\,km/h. The cost function for optimization in \eqref{algo:additional_input} in this phase is given by $L^g_{p1,t}(g;\measx_t,\estx_t,\hat{u}^{\mu,s})= \sum_{k=1}^\mu L_{p1}\big(\hat{x}^g_{k-1},T_k(g+\hat{u}^{\mu,s})+K\hat{x}^g_{k-1};\estx_{y,t},\estx_{\varDelta,t}\big)$, where $\estx_{y,t}$ and $\estx_{\varDelta,t}$ are the estimates of the optimal steady-state position and velocity obtained in \eqref{algo:OGD} at time $t$, $\hat{u}^{\mu,s}:=\begin{bmatrix} \left(\sigma\predu_{t-1}\right)^\top & \left(\ssinput_{t-1}\right)^\top \end{bmatrix}^\top$, and $\hat{x}^g_k\in\mathbb R^n$ is the state at time $k$ resulting from application of $g+\hat{u}^{\mu,s}$ together with the stabilizing feedback and starting from $x_0^g = \measx_t$.
	\item When the controlled car comes close to the slower moving vehicle in front (i.e., $p_{x,t}^c-p_{x,t}\leq100$\,m, where $p_{x,t}^c$ is the longitudinal position of the slower moving vehicle at time $t$), it is detected by the online planner. In this phase, the planner decides to stay behind the vehicle in front. For that, we use a constant cost function $L_{p2}(x,u;\theta_{p2}^y,\theta_{p2,t}^\varDelta)=\frac{1}{2}\norm{p_y-\theta_{p2}^y}^2 + \frac{1}{2} \norm{\varDelta-\theta_{p2,t}^\varDelta}^2 + 50 \norm{u}^2$, where $\theta_{p2}^y = 0$\,m. The optimal velocity $\theta_{p2,t}^\varDelta$ in this phase is calculated by estimating the velocity of the slower vehicle ahead $\varDelta^c$ using the current measured distance $\tilde d_t = d_t + v^d_t$ (where $v^d_t$ is the measurement noise of the sensor measuring the distance between the cars) and the previously measured distance $\tilde d_{t-1}$ as $\varDelta^c \approx \tilde \varDelta^c_t = \frac{\tilde d_t - \tilde d_{t-1}}{\tau} + \tilde \varDelta_t$, where $\tilde\varDelta_t$ is the measured velocity at time $t$. The estimated value is then set as the desired velocity $\theta_{p2,t}^\varDelta = \tilde \varDelta^c_t$. Furthermore, we include a soft constraint to enforce a safety distance of $50$\,m to the slower vehicle ahead given by $\hat{d}_k \geq 50 - \epsilon$ for all $k\in\mathbb{N}_{[0,\mu]}$, where $\hat{d}_{k}$ is the predicted distance to the slower vehicle ahead when applying $g+\hat{u}^{\mu,s}$. The predicted distance $\hat{d}_k$ is computed based on the measured distance $\tilde{d}_t$ under the assumption that the slower vehicle moves with constant speed $\tilde{\varDelta}^c_t$. Then, we solve~\eqref{algo:additional_input} by optimization using the cost function $L^g_{p2,t}(g,\epsilon;\measx_t,\estx_t,\hat{u}^{\mu,s})=100\epsilon^2 + \sum_{k=1}^\mu L_{p2}\big( \hat{x}^g_{k-1}, T_k(g+\hat{u}^{\mu,s})+K\hat{x}^g_{k-1}; \estx_{y,t},\estx_{\varDelta,t} \big)$.
	\item At $t=20$\,s, the online planner decides to overtake the slow vehicle in front. Therefore, the cost function is switched again to $L_{p3}(x,u;\theta_{p3}^y,\theta_{p3}^\varDelta)=\frac{1}{2}\norm{p_y-\theta_{p3}^y}^2 + \frac{5}{2} \norm{\varDelta-\theta_{p3}^\varDelta}^2 + 50 \norm{u}^2$, where $\theta^y_{p3}=3$\,m and $\theta_{p3}^\varDelta = 130$\,km/h, i.e., the controlled car shall move to the left lane and accelerate. The additional weighting $\frac{5}{2}$ (compared to the first phase) on the term penalizing the velocity encourages rapid acceleration, so that the controlled car overtakes the slower vehicle quickly. Note that $\theta_{p3}^\varDelta$ is on the boundary of the constraints. We solve the optimization problem in \eqref{algo:additional_input} with $L^g_{p3,t}(g;\measx_t,\estx_t,\hat{u}^{\mu,s})= \sum_{k=1}^\mu L_{p3}\big(\hat{x}^g_{k-1},T_k(g+\hat{u}^{\mu,s})+K\hat{x}^g_{k-1};\estx_{y,t},\estx_{\varDelta,t}\big)$.
\end{enumerate}

\begin{figure}
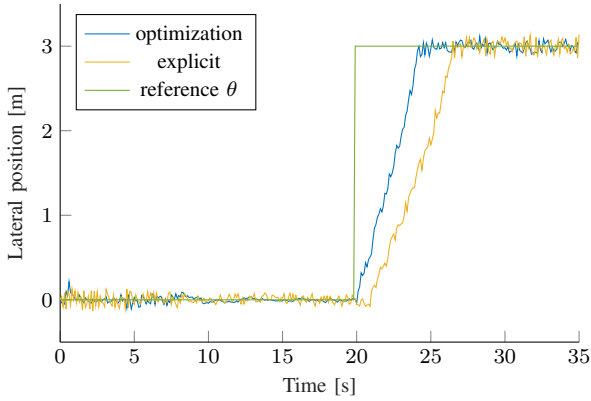

	\centering
	\footnotesize
	\setlength{\hohe}{4.5cm}
	\setlength{\breite}{.4\textwidth}
	\include{figures/lateral_position}
	\vspace{-.75cm}
	\caption{Lateral position of the controlled car in closed loop for two variants of Algorithm~\ref{alg} (optimization based solution of \eqref{algo:additional_input} (blue) and the explicit solution of \eqref{algo:additional_input} (yellow)) and reference position (green).}
	\label{fig:lat_pos}
\end{figure}
\begin{figure}
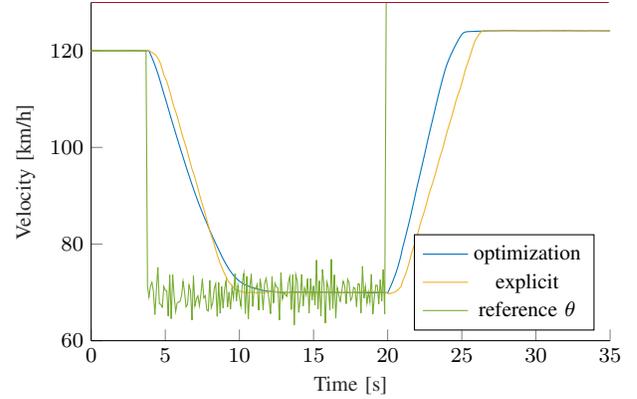

	\centering
	\footnotesize
	\setlength{\hohe}{4.5cm}
	\setlength{\breite}{.4\textwidth}
	\include{figures/velocity}
	\vspace{-.75cm}
	\caption{Velocity of the controlled car for two variants of Algorithm~\ref{alg} (optimization based solution of \eqref{algo:additional_input} (blue) and the explicit solution of \eqref{algo:additional_input} (yellow)), reference velocity (green) and constraint (red).}
	\label{fig:velocity}
\end{figure}

\begin{figure}
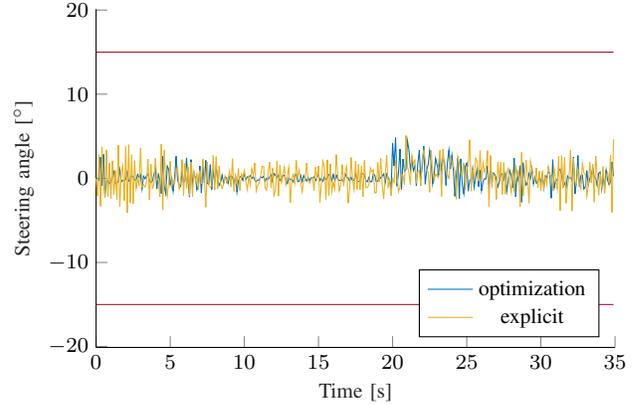

	\centering
	\footnotesize
	\setlength{\hohe}{4.5cm}
	\setlength{\breite}{.4\textwidth}
	\include{figures/angle}
	\vspace{-.75cm}
	\caption{Steering angle of the controlled car in closed loop for two variants of Algorithm~\ref{alg} (optimization based solution of \eqref{algo:additional_input} (blue) and the explicit solution of \eqref{algo:additional_input}) and constraints (red).}
	\label{fig:angle}
\end{figure}

\begin{figure}
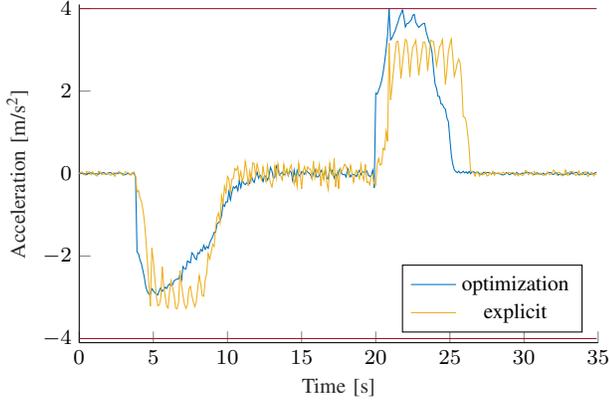

	\centering
	\footnotesize
	\setlength{\hohe}{4.5cm}
	\setlength{\breite}{.4\textwidth}
	\include{figures/acceleration}
	\vspace{-.75cm}
	\caption{Acceleration of the controlled car for two variants of Algorithm~\ref{alg} (optimization based solution of \eqref{algo:additional_input} (blue) and the explicit solution of \eqref{algo:additional_input} (yellow)) and constraints (red).}
	\label{fig:accel}
\end{figure}

The scenario described above is shown in Figure~\ref{fig:scenario} together with the closed-loop trajectories. Algorithm~1 stays on the correct lane and keeps a safety distance of approximately $55$\,m to the slower vehicle ahead. Furthermore, the car's lateral position, velocity and control inputs are shown in Figures \ref{fig:lat_pos}--\ref{fig:accel}. The variation in the reference velocity in Figure~\ref{fig:velocity} is due to the noisy estimation of the slower vehicle's velocity in the second phase. In all figures, Algorithm 1 is able to robustly satisfy all constraints while achieving good reference tracking. In particular, the optimization-based variation of Algorithm~\ref{alg} fully exploits the feasible range of control inputs as can be seen in Figure~\ref{fig:accel}. As noted before and can be seen in Figure \ref{fig:velocity}, the velocity reference in Phase 3, $\theta_{p3}^\varDelta = 130$\,km/h, is on the boundary of the constraints. Due to the constraint tightening approach, the controlled car cannot reach this velocity in steady state and accelerates only to approximately $124$\,km/h.


In a second simulation, we additionally simulate a variant of Algorithm~\ref{alg} that implements the explicit solution of \eqref{algo:additional_input} instead of optimizing the input sequence $g_t$ with respect to the cost functions $L^g$ defined above. All other parameters of Algorithm~\ref{alg} and the simulation remain the same. The results are shown in Figures \ref{fig:scenario}-\ref{fig:accel} together with the closed-loop trajectories of the first simulation. As can be seen, this simpler variant of Algorithm~\ref{alg} performs almost exactly equal to the first variant in the first phase of the simulation. However, in the second phase, this variant of Algorithm~\ref{alg} switches more frequently between braking and accelerating as can be seen in Figure~\ref{fig:accel}. Moreover, in the last phase, the controlled car is slow to move to the left lane when overtaking the slower vehicle in front. This behavior is caused by the explicit solution of \eqref{algo:additional_input}: The explicit solution minimizes the additional control effort $\norm{g_t}$, which leads to higher sensitivity with respect to changes of $\theta^\varDelta_{p2,t}$, i.e., the noisy estimate of the slower vehicle's velocity. In conclusion, choosing a cost function in \eqref{algo:additional_input} that specifies desirable transient behavior for the controlled system yields improved performance in this numerical case study.

\section{Conclusion} \label{sec:conclusion}

In this paper, we propose an algorithm for controlling linear dynamical systems subject to time-varying and a priori unknown cost functions, state and input constraints, exogenous disturbances, and measurement noise. This algorithm is based on the online convex optimization framework and makes use of techniques originally developed in the context of robust model predictive control. In particular, we develop a constraint tightening that ensures robust constraint satisfaction despite the disturbances acting on the system. The proposed algorithm's dynamic regret is shown to be bounded linearly in the variation of the cost functions and the magnitude of the disturbances.

Future work includes generalizing the presented results to more general system classes, e.g., nonlinear systems or time-varying linear systems. Furthermore, analyzing the dynamic regret of the proposed algorithm with respect to the optimal solution of~\eqref{eq:OCP} as a benchmark (as opposed to the sequence of optimal steady states in this work), and possibly modifying the algorithm to achieve a small regret bound, are additional interesting directions for future research.

\appendix
\subsection{Proof of Lemma \ref{lem:feasibility}} 
\label{sec:proofA}
\begin{proof}
	We prove the claims (i)-(iv) by induction. For the base case, note that for $t=0$, the initialization given in Algorithm~\ref{alg} skips the optimization problems \eqref{algo:additional_input}-\eqref{algo:beta}, i.e., we can define w.l.o.g. $g_0=\controlMatrix^\top\left(\controlMatrix\controlMatrix^\top\right)^{-1}(\estx_0-\predx_0)$ and $\beta_0=0$ as feasible solutions of the optimization problems \eqref{algo:additional_input}-\eqref{algo:beta} at time $t=0$, and $\begin{bmatrix} \left(\sigma\predu_{-1}\right)^\top & \left(\ssinput_{-1}\right)^\top \end{bmatrix}^\top=\predu_0$. Furthermore, $\predu_0\in\feasiblesequence(\measx_0)$ holds by Assumption~\ref{ass:init}, and the initialization satisfies $\ssinput_0=\estu_0$, and, thus, $\ssstate_0=G_K\ssinput_0=G_K\estu_0=\estx_0$, i.e., $\ssz_0=\estz_0\in\stronglyfeasibleSS$ again by Assumption~\ref{ass:init}. Finally, using the algorithm's initialization, we obtain
	\begin{align}
		\predx_1 &\refeq{\eqref{algo:prediction}} A_K^\mu \measx_{1} + \controlMatrix \begin{bmatrix} \sigma \predu_{0} \\ \ssinput_{0} \end{bmatrix} \nonumber \\ 
		&\mkern-12mu\refeq{\eqref{eq:meas_state_dynamics},\eqref{algo:control_input}} A_K^{\mu} \left( A_K \measx_{0} + B T_1 \predu_{0} + \noise_{1}\right) + \controlMatrix \begin{bmatrix} \sigma \predu_{0} \\ \ssinput_{0} \end{bmatrix} \nonumber \\		
		&= A_K \left(A_K^\mu \measx_0 + \controlMatrix\predu_0 \right) + B\ssinput_0 + A_K^\mu \noise_1 \nonumber \\
		&= A_K \left( A_K^\mu \estx_0 + \controlMatrix \left( \one_\mu \otimes \estu_0 \right) \right) + B\estu_0  + A_K^\mu \noise_1 \nonumber \\
		&= \estx_0 + A_K^\mu \noise_1  = \ssstate_0 + A_K^\mu \noise_1, \label{eq:predu_init}
	\end{align}
	which implies $\predx_1\in\{\ssstate_0\}\oplus\RPI_\mu^*$ by definition of $\RPI_\mu^*$.

	Hence, for the induction step, we fix any $t\in\mathbb{N}$ and assume that  (i)-(iv) are satisfied for $[0,t]\subseteq\mathbb{N}$.
	
	First, we show that these assumptions imply existence of a feasible solution to~\eqref{algo:additional_input}-\eqref{algo:beta} at time~$t+1$, $\predu_{t+1}\in\feasiblesequence(\measx_{t+1})$, and $\begin{bmatrix} (\sigma\predu_{t})^\top & (\ssinput_{t})^\top \end{bmatrix}^\top \in\feasiblesequence(\measx_{t+1})$, i.e., satisfaction of (i) and (ii) at time~$t+1$. To do so, let $\hat{u}^{\mu,c}_{t+1} := \begin{bmatrix} (\sigma\predu_{t})^\top & (\ssinput_{t})^\top \end{bmatrix}^\top$. In order to show that $\hat{u}^{\mu,c}_{t+1}\in\feasiblesequence(\measx_{t+1})$, we verify below the constraints in~\eqref{eq:tightened_constraint_set}. Note that $\hat{u}^{\mu,c}_{t+1}\in\feasiblesequence(\measx_{t+1})$ would imply that $g^c_{t+1}=\controlMatrix^\top\left(\controlMatrix\controlMatrix^\top\right)^{-1}(\estx_{t+1}-\predx_{t+1})$ and $\beta^c_{t+1}=0$ are feasible candidate solutions to~\eqref{algo:additional_input}-\eqref{algo:beta} at time~$t+1$, because $c_g\geq\norm{\controlMatrix^\top\left(\controlMatrix\controlMatrix^\top\right)^{-1}}$, i.e., satisfaction of (i) at time $t+1$. Furthermore, we note that existence of a feasible candidate solution to~\eqref{algo:additional_input}-\eqref{algo:beta} at time~$t+1$ would imply $\predu_{t+1}\in\feasiblesequence(\measx_{t+1})$. Thus, $\hat{u}^{\mu,c}_{t+1}\in\feasiblesequence(\measx_{t+1})$ would imply satisfaction of (i) and (ii) at time $t+1$ as desired. To prove $\hat{u}^{\mu,c}_{t+1}\in\feasiblesequence(\measx_{t+1})$, we fix any\footnote{If $\mu=1$, then \eqref{eq:mu=1_1} and \eqref{eq:mu=1_2} are sufficient to obtain the desired result.} $\tau\in[0,\mu-2]$. Then, we have
	\begin{align*}
		&A_K^{\tau+1} \measx_{t+1} + \sum_{i=0}^\tau A_K^i B T_{\tau-i+1} \hat{u}^{\mu,c}_{t+1} \\
		\refeq{\eqref{eq:meas_state_dynamics},\eqref{algo:control_input}} &A_K^{\tau+1} (A_K \measx_{t} + BT_1 \predu_{t} + \noise_{t+1}) + \sum_{i=0}^\tau A_K^i B T_{\tau-i+2} \predu_{t} \\
		=~ &A_K^{\tau+2} \measx_{t} + \sum_{i=0}^{\tau+1} A_K^i B T_{\tau-i+2} \predu_{t} + A_K^{\tau+1} \noise_{t+1} \\
		\in~ &\constraintX \ominus \sum_{j=0}^{\tau+1} A_K^j \setW \oplus A_K^{\tau+1} \setW
		\subseteq~ \constraintX \ominus \sum_{j=0}^{\tau} A_K^j \setW,
	\end{align*}
	where the last line follows from $\predu_{t}\in\feasiblesequence(\measx_{t})$. Moreover, due to the induction hypothesis, we have $(\ssstate_{t},\ssinput_t)\in\stronglyfeasibleSS$ and $\predx_{t+1}\in\{\ssstate_t\}\oplus\RPI^*_\mu$. Combining this with $\RPI_\mu^* \oplus \sum_{j=0}^{\mu-1} A_K^j \setW \subseteq \RPI$ (compare~\eqref{eq:RPI_subset}) implies
	\begin{align}
		&A_K^{\mu} \measx_{t+1} + \sum_{i=0}^{\mu-1} A_K^i B T_{\mu-i} \hat{u}^{\mu,c}_{t+1} = A_K^\mu \measx_{t+1} + \controlMatrix \begin{bmatrix} \sigma \predu_{t} \\ \ssinput_{t} \end{bmatrix} \nonumber \\
		\refeq{\eqref{algo:prediction}} &\predx_{t+1} \in \{\ssstate_{t}\} \oplus \RPI_\mu^* \subseteq \constraintX \ominus \RPI \oplus \RPI_\mu^* \nonumber \\
		&\phantom{\predx_{t+1}}\subseteq \constraintX \ominus \sum_{j=0}^{\mu-1} A_K^j \setW, \label{eq:mu=1_1}
	\end{align}	
	i.e., the state constraints in~\eqref{eq:tightened_constraint_set} are satisfied.
	Furthermore, for any $\tau\in[0,\mu-2]$,
	\begin{align*}
		&T_{\tau+1} \hat{u}^{\mu,c}_{t+1} + K \left( A_K^{\tau} \measx_{t+1} + \sum_{i=0}^{\tau-1} A_K^i B T_{\tau-i} \hat{u}^{\mu,c}_{t+1} \right) \\
		\refeq{\eqref{eq:meas_state_dynamics},\eqref{algo:control_input}}\, &T_{\tau+1} \begin{bmatrix} \sigma\predu_{t} \\ \ssinput_{t} \end{bmatrix} + K \Bigg( A_K^{\tau+1} \measx_{t} + A_K^\tau B T_1 \predu_{t} \nonumber \\
		&\quad + \sum_{i=0}^{\tau-1} A_K^i B T_{\tau-i+1} \predu_{t} \Bigg) + K A_K^\tau \noise_{t+1} \\
		=~&T_{\tau+2} \predu_{t} + K \left( A_K^{\tau+1} \measx_{t} + \sum_{i=0}^{\tau} A_K^i B T_{\tau-i+1} \predu_{t} \right) \nonumber \\
		&\quad+ K A_K^\tau \noise_{t+1}\\
		\in~ &\constraintU \ominus K \sum_{j=0}^{\tau} A_K^{j}\setW \oplus KA_K^\tau \setW
		\subseteq~ \constraintU \ominus K\sum_{j=0}^{\tau-1} A_K^{j}\setW,
	\end{align*}
	again because $\predu_{t}\in\feasiblesequence(\measx_{t})$, i.e., the induction hypothesis. Finally, similar arguments yield
	\begin{align*}
		\tilde u_t := &T_\mu \hat{u}^{\mu,c}_{t+1} + K \left( A_K^{\mu-1} \measx_{t+1} + \sum_{i=0}^{\mu-2} A_K^i B T_{\mu-i-1} \hat{u}^{\mu,c}_{t+1} \right) \\
		\refeq{\eqref{eq:meas_state_dynamics},\eqref{algo:control_input}}&\ssinput_{t} + K \left( A_K^{\mu} \measx_{t} + \controlMatrix \predu_{t} \right)+ K A_K^{\mu-1} \noise_{t+1} \\
		\refeq{\eqref{algo:predicted_input}}~ &\ssinput_{t} + K \left( A_K^{\mu} \measx_{t} + \controlMatrix \begin{bmatrix} \sigma\predu_{t-1} \\ \ssinput_{t-1} \end{bmatrix} + \controlMatrix \beta_t g_t \right) \\
		\quad &+ K A_K^{\mu-1} \noise_{t+1} \\
		\refeq{\eqref{algo:prediction},\eqref{algo:additional_input}} &\ssinput_t + (1-\beta_t) K\predx_t + \beta_t K \estx_t + K A_K^{\mu-1} \noise_{t+1} \\
		\refeq{\eqref{algo:us_t}} ~& (1-\beta_t) (\ssinput_{t-1} + K\ssstate_{t-1}) + \beta_t (\estu_{t}+K\estx_t) \\
		&\quad+ (1-\beta_t)K(\predx_t - \ssstate_{t-1}) + KA_K^{\mu-1} \noise_{t+1}.
	\end{align*}	
	for any $t\in\mathbb{N}_{\geq1}$. Thus, from the induction hypothesis $(\ssstate_{t-1},\ssinput_{t-1})\in\stronglyfeasibleSS$ and $\predx_{t}-\ssstate_{t-1}\in\RPI_\mu^*$, $(\estx_{t},\estu_{t})\in\stronglyfeasibleSS$ by~\eqref{algo:OGD}, and since the set $\stronglyfeasibleSS$ is convex, we obtain
	\begin{align}
		\tilde u_t &\in~\constraintU \ominus K\RPI \oplus (1-\beta_{t}) K\RPI_\mu^* \oplus KA_K^{\mu-1}\setW \nonumber \\
		&\subseteq \constraintU \ominus K\RPI^* \oplus K\RPI_\mu^* \oplus KA_K^{\mu-1}\setW \nonumber \\
		&\subseteq \constraintU \ominus K \sum_{j=0}^{\mu-2} A_K^j \setW,  \label{eq:mu=1_2}
	\end{align}
	for any $t\in\mathbb{N}_{\geq1}$, where the second line follows from $\RPI^*\subseteq\RPI$, $0\in\RPI_\mu^*$, convexity of $\RPI_\mu^*$, and $0\leq\beta_t\leq1$. Furthermore, in case $t=0$, recalling that $\estx_0=\ssstate_0$, we have
	\begin{align*} 
		\tilde u_0 = &T_\mu \hat{u}^{\mu,c}_1 + K\left(A_K^{\mu-1} \measx_1 + \sum_{i=0}^{\mu-2} A_K^iBT_{\mu-i-1} \hat{u}^{\mu,c}_1 \right) \\
		\refeq{\eqref{eq:meas_state_dynamics},\eqref{algo:control_input}} &\ssinput_{0} + K \left( A_K^{\mu} \measx_{0} + \controlMatrix \predu_{0} \right)+ K A_K^{\mu-1} \noise_{1} \\
		=~&\ssinput_0 + K \left( A_K^\mu \estx_0 + \controlMatrix \left( \one_\mu\otimes \estu_0 \right) \right) + KA_K^{\mu-1}\noise_1 \\
		=~& \ssinput_0 + K\ssstate_0 + KA_K^{\mu-1}\noise_1,
	\end{align*}
	i.e., \eqref{eq:mu=1_2} holds for any $t\in\mathbb{N}$. Thus, the input constraints in~\eqref{eq:tightened_constraint_set} are also satisfied, which proves $\hat{u}^{\mu,c}_{t+1}\in\feasiblesequence(\measx_{t+1})$. As discussed above, the fact that $\hat{u}^{\mu,c}_{t+1}\in\feasiblesequence(\measx_{t+1})$ implies that $\beta^c_{t+1}=0$ together with $g^c_{t+1}=\controlMatrix^\top\left(\controlMatrix\controlMatrix^\top\right)^{-1}\left(\estx_{t+1}-\predx_{t+1}\right)$ is a feasible candidate solution to \eqref{algo:additional_input}-\eqref{algo:beta} at time $t+1$. Thus, $\predu_{t+1}\in\feasiblesequence(\measx_{t+1})$ by \eqref{algo:predicted_input} and the constraint in~\eqref{algo:beta}.
	
	In order to conclude the proof by induction, it remains to show that $\ssz_{t+1}\in\stronglyfeasibleSS$ and $\predx_{t+2}\in\{\ssstate_{t+1}\} \oplus \RPI^*_\mu$, i.e., that (iii) and (iv) are satisfied at time $t+1$. To prove the former, we have that existence of a feasible solution $\beta_{t+1}$ for~\eqref{algo:beta} at time~$t+1$ as shown above implies
	\begin{align}
		\ssstate_{t+1} &:= G_K\ssinput_{t+1} \refeq{\eqref{algo:us_t}} (1-\beta_{t+1})G_K\ssinput_{t} + \beta_{t+1} G_K \estu_{t+1} \nonumber \\
		&= (1-\beta_{t+1}) \ssstate_{t} + \beta_{t+1} \estx_{t+1}. \label{eq:xs_convex_comb}
	\end{align}	
	Combining this with \eqref{algo:us_t} yields
	\begin{equation}
		\ssz_{t+1}=(1-\beta_{t+1}) \ssz_{t} + \beta_{t+1} \estz_{t+1}. \label{eq:zs_convex_comb}
	\end{equation}
	Thus, $\ssz_{t+1}\in\stronglyfeasibleSS$ holds due to ´the induction hypothesis $\ssz_t\in\stronglyfeasibleSS$, $\estz_{t+1}\in\stronglyfeasibleSS$ by definition~\eqref{algo:OGD}, and convexity of $\stronglyfeasibleSS$.
	
	It remains to prove $\predx_{t+2}\in\{\ssstate_{t+1}\}\oplus\RPI_\mu^*$. By similar arguments as in~\eqref{eq:predu_init}, we get
	\begin{align*}
		&\predx_{t+2} \refeq{\eqref{algo:prediction}} A_K^\mu \measx_{t+2} + \controlMatrix \begin{bmatrix} \sigma \predu_{t+1} \\ \ssinput_{t+1} \end{bmatrix}  \\ 
		\refeq{\eqref{eq:meas_state_dynamics},\eqref{algo:control_input}} &A_K^{\mu} \left( A_K \measx_{t+1} + B T_1 \predu_{t+1} + \noise_{t+2}\right) + \controlMatrix \begin{bmatrix} \sigma \predu_{t+1} \\ \ssinput_{t+1} \end{bmatrix}  \\
		=\mkern13mu&A_K \left( A_K^{\mu} \measx_{t+1} + \controlMatrix \predu_{t+1} \right) + B\ssinput_{t+1} + A_K^\mu \noise_{t+2}  \\
		\refeq{\eqref{algo:predicted_input}}\mkern10mu &A_K \left( A_K^\mu \measx_{t+1} + \controlMatrix \begin{bmatrix} \sigma \predu_{t} \\ \ssinput_{t} \end{bmatrix} + \beta_{t+1} \controlMatrix g_{t+1} \right)  \\
		&\qquad+ B\ssinput_{t+1} + A_K^\mu \noise_{t+2}  \\
		\refeq{\eqref{algo:prediction},\eqref{algo:additional_input}} &(1{-}\beta_{t+1}) A_K\predx_{t+1} + \beta_{t+1} A_K \estx_{t+1} + B\ssinput_{t+1} + A_K^\mu \noise_{t+2}.
	\end{align*}
	Combining this result with $\ssstate_{t+1} = G_K\ssinput_{t+1}$, which implies $\ssstate_{t+1} = A_K\ssstate_{t+1}+B\ssinput_{t+1}$, we obtain
	\begin{align}
		&\predx_{t+2} - \ssstate_{t+1} = \predx_{t+2} - (A_K\ssstate_{t+1} + B\ssinput_{t+1}) \nonumber \\
		\refeq{\eqref{eq:xs_convex_comb}} &(1-\beta_{t+1}) A_K \left(\predx_{t+1} - \ssstate_{t} \right)+ A_K^\mu \noise_{t+2}. \label{eq:pred_ss_error_rec}
	\end{align}
	Thus, $\predx_{t+2}-\ssstate_{t+1}\in A_K\RPI_\mu^*\oplus A_K^\mu\setW\subseteq \RPI_\mu^*$ since $0\leq\beta_{t+1}\leq1$ and by convexity and the definition of the RPI set $\RPI_\mu^*$. Hence, the proof by induction is complete and we have shown that (i)-(iv) hold for all $t\in\mathbb{N}$.
	
	It remains to show (v) that the constraints $x_t\in\constraintX$ and $u_t\in\constraintU$ are satisfied for all $t\in\mathbb{N}$. Since $\predu_t \in \feasiblesequence(\measx_t)$ for all $t\in\mathbb{N}$, we have
	\[
		u_t = T_1 \predu_t + K\measx_t \in \constraintU
	\]
	and
	\begin{align*}
		x_{t+1} &= A x_t + B u_t + w_t = A_K \tilde x_t + BT_1 \predu_t - Av_t + w_t \\
		&\in \constraintX \ominus \setW \oplus (-A\mathcal V) \oplus \mathcal W \subseteq \constraintX,
	\end{align*}
	for all $t\in\mathbb{N}$ by~\eqref{eq:tightened_constraint_set}, which concludes the proof since $x_0\in\constraintX$ by Assumption \ref{ass:init}.
\end{proof}

\subsection{Proof of Lemma \ref{lem:mincontr}} \label{sec:proofB}

\begin{proof}		
	First, note that there exists $\delta>0$ such that $u+Kx\in\constraintU\ominus K\RPI \ominus \delta \ball_m$ and $x\in\constraintX\ominus\RPI\ominus\delta\ball_n$ hold for all $(x,u)\in\stronglyfeasibleSS\subseteq \text{rel\,int } \feasibleSS$. Moreover, since $A_K$ is Schur stable, there exist $c_A\geq1$ and $\phi\in[0,1)$ such that $\norm{A_K^t}\leq c_A\phi^t$ holds for all $t\in\mathbb{N}$. Recall the diameter of the sets $\constraintX$ and $\constraintU$ are $d_\constraintX$ and $d_\constraintU$, respectively. Finally, in order to shorten notation we let $c_x := c_A\frac{1-\phi^\mu}{1-\phi}\norm{B}c_gd_\constraintX$, $c_u := \max\left(d_\constraintU+\norm{K}d_\constraintX,c_gd_\constraintX\right)$, $c_{ux} := c_x+\mu\norm{B}\left(1+c_u\right)$, and $c_\mathrm{max} := \max \left(c_A c_{ux},\norm{K}c_Ac_{ux}+1+c_u \right)$. Fix any $t\in\mathbb{N}_{\geq1}$. We proceed by a case distinction.
			
	\noindent\underline{\textit{Case 1:}} $\sum_{k=0}^{\mu-1} \beta_{t+k} > \delta_\beta :=  \min\left(\frac{\delta}{c_\mathrm{max}},c_gd_\constraintX\right) > 0$. In this case, there exists $k^*\in\mathbb{N}_{[0,\mu-1]}$ such that $1\geq\beta_{t+k^*} > \frac{\delta_\beta}{\mu}$. Hence,
	\[
		\prod_{k=0}^\mu (1+\beta_{t+k}) \leq 1-\beta_{t+k^*} < 1-\frac{\delta_\beta}{\mu} < 1.
	\]
	
	\noindent\underline{\textit{Case 2:}} $\sum_{k=0}^{\mu-1} \beta_{t+k} \leq \delta_\beta$. Intuitively, in this case, $\measx_{t+\mu}$ is close to $\ssstate_{t-1}$, and, similarly, $T_j\begin{bmatrix} (\sigma\predu_{t+\mu-1})^\top&(\ssinput_{t+\mu-1})^\top\end{bmatrix}^\top$ is close to $\ssinput_{t-1}$ for all $j\in\mathbb{N}_{[1,\mu]}$, because $\beta_{t+k}$ being small for all $k\in\mathbb{N}_{[0,\mu-1]}$ implies that the predicted input sequence $\predu_{t-1}$ has been applied over $\mu$ time steps with only small modifications, compare~\eqref{algo:predicted_input}. In the following, we formalize this intuition. Using \eqref{eq:meas_state_dynamics}, \eqref{algo:predicted_input} and \eqref{algo:control_input}, we get
	\begin{align*}
		&\measx_{t+\mu} \refeq{\eqref{eq:meas_state_dynamics},\eqref{algo:control_input}} A_K\measx_{t+\mu-1} + BT_{1}\predu_{t+\mu-1} + \noise_{t+\mu} \\
		\refeq{\eqref{algo:predicted_input}} &A_K\measx_{t+\mu-1}{+}BT_{2} \predu_{t+\mu-2} + \beta_{t+\mu-1} B T_1 g_{t+\mu-1} + \noise_{t+\mu}.
	\end{align*}
	Using this equation recursively yields
	\begin{align}
		\measx_{t+\mu}&=A_K^2 \measx_{t+\mu-2} + A_K BT_{2} \predu_{t+\mu-3} + BT_{2} \predu_{t+\mu-2} \nonumber \\
		&\quad + \beta_{t+\mu-1} BT_1 g_{t+\mu-1} + \beta_{t+\mu-2} A_K B T_1 g_{t+\mu-2} \nonumber \\
		&\quad + \noise_{t+\mu} + A_K \noise_{t+\mu-1} \nonumber \\
		&\mkern-3mu\refeq{\eqref{algo:predicted_input}} A_K^2 \measx_{t+\mu-2} + A_K BT_{2} \predu_{t+\mu-3} + BT_{3} \predu_{t+\mu-3} \nonumber \\
		&\quad  + \beta_{t+\mu-2} \left( BT_{2}g_{t+\mu-2} + A_K BT_1 g_{t+\mu-2} \right) \nonumber \\
		&\quad + \beta_{t+\mu-1} BT_1 g_{t+\mu-1}+ \noise_{t+\mu} + A_K \noise_{t+\mu-1} \nonumber \\		
		&\,= A_K^\mu \measx_{t} + \controlMatrix \begin{bmatrix} \sigma\predu_{t-1} \\ \ssinput_{t-1} \end{bmatrix} + \sum_{k=1}^{\mu} A_K^{\mu-k} \noise_{t+k} \nonumber \\
		&\quad + \sum_{k=0}^{\mu-1} \beta_{t+k} \left( \sum_{j=k+1}^{\mu} A_K^{\mu-j}B T_{j-k} g_{t+k} \right) \nonumber \\
		\begin{split} \label{eq:xmeas_rec}
			&\,\refeq{\eqref{algo:prediction}} \predx_{t} + \sum_{k=1}^{\mu} A_K^{\mu-k} \noise_{t+k}  \\
			&\qquad+ \sum_{k=0}^{\mu-1} \beta_{t+k} \left( \sum_{j=k+1}^{\mu} A_K^{\mu-j}B T_{j-k} g_{t+k} \right).
		\end{split}
	\end{align}
	Second, for all $\tau\in[1,\mu]$ we obtain $T_\tau \begin{bmatrix} \sigma \predu_{t+\mu-1} \\ \ssinput_{t+\mu-1} \end{bmatrix} $
	\begin{align*}
		&\mkern-15mu\refeq{\eqref{algo:us_t},\eqref{algo:predicted_input}} T_{\tau} 
		\begin{bmatrix}
		\sigma \left( \begin{bmatrix} \sigma\predu_{t+\mu-2} \\ \ssinput_{t+\mu-2} \end{bmatrix} + \beta_{t+\mu-1} g_{t+\mu-1} \right) \\
		\ssinput_{t+\mu-2} + \beta_{t+\mu-1} \left(\estu_{t+\mu-1}-\ssinput_{t+\mu-2} \right)
		\end{bmatrix} \\
		&=T_{\tau} 
		\left(
		\begin{bmatrix} 
			\sigma\left(\sigma \predu_{t+\mu-2} \right) \\
			\ssinput_{t+\mu-2} \\
			\ssinput_{t+\mu-2}
		\end{bmatrix} + \beta_{t+\mu-1} \begin{bmatrix}
			\sigma g_{t+\mu-1} \\
			\estu_{t+\mu-1}-\ssinput_{t+\mu-2}
		\end{bmatrix} \right),
	\end{align*}
	which implies
	\begin{align}
			T_\tau \begin{bmatrix} \sigma \predu_{t+\mu-1} \\ \ssinput_{t+\mu-1} \end{bmatrix}&\refeq{\eqref{algo:us_t},\eqref{algo:predicted_input}} \ssinput_{t-1} + \sum_{k=0}^{\tau-1} \beta_{t+k} \left(\estu_{t+k} - \ssinput_{t+k-1} \right) \nonumber \\
			&\qquad + \sum_{k=\tau}^{\mu-1} \beta_{t+k}T_{\mu-k+\tau} g_{t+k}. \label{eq:pred_u_rec}
	\end{align}
	Thus, from~\eqref{eq:xmeas_rec}, if $\beta_{t+k}=0$ for all $k\in\mathbb{N}_{[0,\mu-1]}$, we obtain $\measx_{t+\mu} = \predx_{t} + \sum_{k=1}^{\mu}A_K^{\mu-k}\noise_{t+k} \in \{\ssstate_{t-1}\}\oplus\RPI_\mu^*\oplus\sum_{k=0}^{\mu-1}A_K^k\setW \subseteq \{\ssstate_{t-1}\}\oplus\RPI$ by Lemma~\ref{lem:feasibility}, and from~\eqref{eq:pred_u_rec}, we get $T_\tau \begin{bmatrix} (\sigma \predu_{t+\mu-1})^\top & (\ssinput_{t+\mu-1})^\top \end{bmatrix}^\top = \ssinput_{t-1}$ for all $\tau\in\mathbb{N}_{[1,\mu]}$. If $\beta_{t+k}\neq0$, we get additional error terms that depend linearly on $\beta_{t+k}$.
	In the following, we bound these error terms for the case $\sum_{k=0}^{\mu-1} \beta_{t+k}\leq\delta_\beta$. For the error term in~\eqref{eq:xmeas_rec}, we obtain
	\begin{align*}
		\norm{e_x} := &\norm{\sum_{k=0}^{\mu-1} \beta_{t+k} \left( \sum_{j=k+1}^\mu A_K^{\mu-j}B T_{j-k} g_{t+k} \right)}  \\
		\refleq{\eqref{algo:additional_input}} &\left(\sum_{k=0}^{\mu-1} \beta_{t+k} \right) \left( \sum_{j=1}^\mu \norm{A_K^{\mu-j}}\norm{B} c_g d_{\constraintX} \right)  \\
		\leq &\delta_\beta c_A \frac{1-\phi^{\mu}}{1-\phi} \norm{B} c_g d_\constraintX,
	\end{align*}
	which implies
	\begin{equation}
		e_x \in c_x\delta_\beta \ball_n. \label{eq:xmeas_in_ball}
	\end{equation}
	Furthermore, we have $\estz_t,\ssz_t\in\stronglyfeasibleSS$ for all $t\in\mathbb{N}$ by Lemma~\ref{lem:feasibility} and~\eqref{algo:OGD}, which implies $\estu_t+K\estx_t\in\constraintU$ and $\ssinput_t+K\ssstate_t\in\constraintU$ for all $t\in\mathbb{N}$. Thus, for the error terms in~\eqref{eq:pred_u_rec}, we obtain
	\begin{align*}
		\norm{e_u}:=&\left\lVert\sum_{k=0}^{\tau-1} \beta_{t+k} \left(\estu_{t+k}-\ssinput_{t+k-1}\right) \right. \\
		&\quad \left. + \sum_{k=\tau}^{\mu-1} \beta_{t+k} T_{\mu-k+\tau} g_{t+k}\right\rVert  \\
		\leq&\sum_{k=0}^{\tau-1} \beta_{t+k} \norm{(\estu_{t+k}+K\estx_{t+k})-(\ssinput_{t+k-1}+K\ssstate_{t+k-1})} \\
		&\quad + \sum_{k=0}^{\tau-1} \beta_{t+k} \norm{K\estx_{t+k}-K\ssstate_{t+k-1}} \\
		&\quad + \sum_{k=\tau}^{\mu-1} \beta_{t+k}\norm{T_{\mu-k+\tau}g_{t+k}} \\
		\refleq{\eqref{algo:additional_input}}&\sum_{k=0}^{\tau-1} \beta_{t+k} d_\constraintU +\sum_{k=0}^{\tau-1} \beta_{t+k}\norm{K}d_\constraintX + \sum_{k=\tau}^{\mu-1} \beta_{t+k} c_g d_\constraintX \\
		\leq\mkern3mu& \max\left(d_\constraintU+\norm{K}d_\constraintX,c_gd_\constraintX\right)\delta_\beta
	\end{align*}
	for all $\tau\in\mathbb{N}_{[1,\mu]}$. Hence,
	\begin{equation}
		e_u\in c_u\delta_\beta\ball_m. \label{eq:aux_set_relation}
	\end{equation}
	Furthermore, Lemma~\ref{lem:feasibility} shows $\predx_{t}\in\{\ssstate_{t-1}\}\oplus\RPI_\mu^*$. Thus,
	\begin{align}
		\measx_{t+\mu} - \ssstate_{t-1} \refin{\eqref{eq:xmeas_rec},\eqref{eq:xmeas_in_ball}} &\RPI_\mu^* \oplus \sum_{k=0}^{\mu-1}A_K^k\setW \oplus c_x\delta_\beta \ball_n \nonumber \\
		\subseteq \mkern15mu&\RPI \oplus c_x\delta_\beta \ball_n \label{eq:measx-xs_set_relation}
	\end{align}
	holds for all $t\in\mathbb{N}_{\geq1}$.
	
	Having established bounds on the error terms in~\eqref{eq:xmeas_rec} and~\eqref{eq:pred_u_rec}, we make use of a candidate solution to \eqref{algo:beta} at time $t+\mu$ next. This candidate solution is given by $\hat{u}^{\mu,c}_{t+\mu} := \begin{bmatrix} \left(\sigma\predu_{t+\mu-1}\right)^\top & \left(\ssinput_{t+\mu-1}\right)^\top \end{bmatrix}^\top + \beta^c_{t+\mu}g^c_{t+\mu}$, where $g^c_{t+\mu}\in\mathbb{R}^{\mu m}$ is a feasible solution for\footnote{Note that such a feasible solution exists by Lemma~\ref{lem:feasibility}.}~\eqref{algo:additional_input}, and $\beta^c_{t+\mu}\in[0,1]$ is such that $\beta^c_{t+\mu}g^c_{t+\mu} \in \mathcal G_\delta := \{ g \in \mathbb R^{\mu m}: \norm{T_j g}\leq\delta_\beta \quad \forall j\in\mathbb{N}_{[1,\mu]}\}$. Additionally, we denote the state trajectory corresponding to the candidate input $\hat{u}^{\mu,c}_{t+\mu}$ by 
	\[
		\tilde{x}_{\tau+1}^c := A_K^{\tau+1}\measx_{t+\mu} + \sum_{j=0}^{\tau} A_K^{j}BT_{\tau+1-j} \hat{u}^{\mu,c}_{t+\mu}.
	\] 
	We proceed to show that $\hat{u}^{\mu,c}_{t+\mu}\in\feasiblesequence(\measx_{t+\mu})$, i.e., $\beta^c_{t+\mu}$ is a feasible solution to~\eqref{algo:beta} at time $t+\mu$. To this end, we get for any $\tau\in\mathbb{N}_{[0,\mu-1]}$,
	\begin{align*}
		&\tilde{x}_{\tau+1}^c-\ssstate_{t-1} = A_K^{\tau+1} \measx_{t+\mu} - \ssstate_{t-1} \\
		&+ \sum_{j=0}^{\tau} \left( A_K^jBT_{\tau+1-j} \begin{bmatrix} \sigma\predu_{t+\mu-1} \\ \ssinput_{t+\mu-1} \end{bmatrix}+ \beta^c_{t+\mu}A_K^jBT_{\tau+1-j} g^c_{t+\mu} \right) \\
		&\refeq{\eqref{eq:pred_u_rec}}  A_K^{\tau+1} \left(\measx_{t+\mu} - \ssstate_{t-1}\right) - \ssstate_{t-1} + A_K^{\tau+1} \ssstate_{t-1} \\
		&\quad + \sum_{j=0}^{\tau} A_K^j B \ssinput_{t-1} + \sum_{j=0}^{\tau} \beta^c_{t+\mu}A_K^jBT_{\tau+1-j} g^c_{t+\mu} \\
		&\quad +\sum_{j=0}^{\tau} A_K^jB \left( \sum_{k=0}^{\tau-j} \beta_{t+k} (\estu_{t+k} - \ssinput_{t+k-1}) \right. \\
		&\qquad + \left. \sum_{k=\tau-j+1}^{\mu-1} \beta_{t+k} T_{\mu-k+\tau+1-j} g_{t+k} \right).
	\end{align*}
	Note that $A_K^{\tau+1} \ssstate_{t-1} + \sum_{i=0}^{\tau} A_K^i B \ssinput_{t-1} = \ssstate_{t-1}$ because $(\ssstate_{t-1},\ssinput_{t-1})$ is a steady state by definition. Thus, recalling the constants $c_A > 0$ and $\phi\in[0,1)$ such that $\norm{A_K^k} \leq c_A\phi^k\leq c_A$ for all $k\in\mathbb{N}$, $\tau\leq\mu-1$, and noting that $A_K^{\tau+1}\RPI \subseteq \RPI \ominus \sum_{j=0}^{\tau} A_K^j \setW$ because $\RPI$ is an RPI set, yields
	\begin{align}
		&\tilde{x}_{\tau+1}^c-\ssstate_{t-1}\refin{\eqref{eq:aux_set_relation},\eqref{eq:measx-xs_set_relation}} A_K^{\tau+1} \RPI \oplus \left(\norm{A_K^{\tau+1}} c_x \delta_\beta \right) \ball_n \nonumber \\
		&\quad \oplus \left(\sum_{j=0}^{\tau} \norm{A_K^j} \norm{B} \delta_\beta \right) \ball_n \oplus \left(\sum_{j=0}^{\tau} \norm{A_K^j} \norm{B} c_u \delta_\beta \right) \ball_n \nonumber \\
		&\subseteq \RPI \ominus \sum_{j=0}^{\tau} A_K^j \setW \oplus c_A c_{ux} \delta_\beta \ball_n \label{eq:xc-xs_t-mu_set_relation}
	\end{align}
	for all $\tau\in\mathbb{N}_{[0,\mu-1]}$. Next, we verify that the constraints in~\eqref{eq:tightened_constraint_set} are satisfied for $\hat{u}^{\mu,c}_{t+\mu}$. For the state constraints, noting that $(\ssstate_{t},\ssinput_{t})\in\stronglyfeasibleSS$ implies $\ssstate_{t}\in\constraintX\ominus\RPI\ominus\delta\ball_n$ as discussed above, we obtain for any $\tau\in\mathbb{N}_{[0,\mu-1]}$,
	\begin{align*}
		\tilde{x}_{\tau+1}^c &= \ssstate_{t} + \left( \tilde{x}_{\tau+1}^c - \ssstate_{t} \right) \refin{\eqref{eq:xc-xs_t-mu_set_relation}} \constraintX \ominus \RPI \ominus \delta\ball_n \oplus \RPI \\
		&\quad\ominus \sum_{k=0}^{\tau} A_K^k\setW \oplus c_Ac_{ux}\delta_\beta \ball_n 
		\subseteq \constraintX \ominus \sum_{j=0}^\tau A_K^j \setW,
	\end{align*}
	because $c_A c_{ux}\delta_\beta \leq \frac{\delta}{c_\mathrm{max}} c_\mathrm{max} = \delta$. Finally, for the input constraints in~\eqref{eq:tightened_constraint_set}, we obtain for any $\tau\in\mathbb{N}_{[0,\mu-1]}$,
	\begin{align*}
		&K\tilde{x}_{\tau}^c + T_{\tau+1}\hat{u}^{\mu,c}_{t+\mu} = K\tilde{x}_{\tau}^c + T_{\tau+1} \begin{bmatrix} \sigma \predu_{t+\mu-1} \\ \ssinput_{t+\mu-1} \end{bmatrix} \\
		&\quad+ \beta^c_{t+\mu} T_{\tau+1} g_{t+\mu}^c \\
		&\refeq{\eqref{eq:pred_u_rec}} K\ssstate_{t-1} + \ssinput_{t-1} + K(\tilde{x}_{\tau}^c - \ssstate_{t-1}) + \beta^c_{t+\mu} T_{\tau+1} g_{t+\mu}^c \\
		&\quad+ \sum_{k=0}^{\tau} \beta_{t+k} \left(\estu_{t+k} - \ssinput_{t+k-1} \right) + \sum_{k=\tau+1}^{\mu-1} \beta_{t+k} T_{\mu-k+\tau+1}g_{t+k} \\
		&\refin{\eqref{eq:aux_set_relation},\eqref{eq:xc-xs_t-mu_set_relation}}\constraintU \ominus K\RPI \ominus \delta \ball_m \oplus K \RPI\ominus K\sum_{k=0}^{\tau-1} A_K^k \setW \\
		&\qquad \oplus \norm{K} c_A c_{ux} \delta_\beta \ball_m \oplus \delta_\beta \ball_m \oplus c_u\delta_\beta \ball_m \\
		&\subseteq~ \constraintU \ominus K\sum_{k=0}^{\tau-1} A_K^k \setW,
	\end{align*}
	because $\delta_\beta (\norm{K}c_Ac_{ux}+1+c_u) \leq \frac{\delta}{c_\mathrm{max}} c_\mathrm{max} = \delta$. 
	
	Summarizing the above, we have shown that, if $\sum_{k=0}^{\mu-1}\beta_{t+k}\leq\delta_\beta$, then 
	\[
		\hat{u}^{\mu,c}_{t+\mu} = \begin{bmatrix} \sigma\predu_{t+\mu-1} \\ \ssinput_{t+\mu-1} \end{bmatrix} + \beta^c_{t+\mu}g^c_{t+\mu} \in \feasiblesequence(\measx_{t+\mu})
	\]
	for any feasible solution $g^c_{t+\mu}$ for~\eqref{algo:additional_input} and all $\beta^c_{t+\mu}\in[0,1]$ that satisfy $\beta^c_{t+\mu} g_{t+\mu}^c \in \mathcal G_\delta$. We prove the required bound on $\prod_{k=0}^{\mu} (1-\beta_{t+k})$ by constructing a candidate solution satisfying $\beta^c_{t+\mu}\geq b$ and $b>0$ such that $\beta^c_{t+\mu} g_{t+\mu}^c \in \mathcal G_\delta$ holds for any $g^c_{t+\mu}$. To this end, we use another case distinction.
	
	\noindent\underline{\textit{Case 2.1:}} $\norm{\estx_{t+\mu}-\predx_{t+\mu}} <\frac{\delta_\beta}{c_g}$. Since $g^c_{t+\mu}$ is a feasible solution to~\eqref{algo:additional_input} at time $t+\mu$, $\norm{g^c_{t+\mu}}\leq c_g\norm{\estx_{t+\mu}-\predx_{t+\mu}}$ holds. Then, we can choose $\beta_{t+\mu}^c = 1$, because $\norm{T_\tau \beta^c_{t+\mu} g^c_{t+\mu}} \leq c_g \norm{\estx_{t+\mu}-\predx_{t+\mu}} < \delta_\beta$ holds for all $\tau\in\mathbb{N}_{[1,\mu]}$. Since $\beta_{t+\mu}\in[0,1]$ is maximized in~\eqref{algo:beta}, it follows that $\beta_{t+\mu}=\beta^c_{t+\mu}=1$. Hence, we obtain
	\begin{equation*}
		\prod_{k=0}^\mu \left( 1- \beta_{t+k}\right) \leq 1-\beta_{t+\mu} = 0.
	\end{equation*}
	
	\noindent \underline{\textit{Case 2.2:}} $\norm{\estx_{t+\mu}-\predx_{t+\mu}} \geq \frac{\delta_\beta}{c_g}$. In this case, we obtain $\beta_{t+\mu} \geq \beta_{t+\mu}^c := \frac{\delta_\beta}{c_gd_\constraintX}$ because $\frac{\delta_\beta}{c_gd_\constraintX} \leq  1$ by definition of $\delta_\beta$, and $\norm{T_\tau \beta^c_{t+\mu} g^c_{t+\mu}} \leq \frac{\delta_\beta}{d_\constraintX} \norm{\estx_{t+\mu}-\predx_{t+\mu}} \leq \delta_\beta$. Thus,
	\[
		\prod_{k=0}^\mu (1-\beta_{t+k}) \leq 1-\beta_{t+\mu} \leq 1-\beta_{t+\mu}^c = 1-\frac{\delta_\beta}{c_gd_\constraintX}<1.
	\]
		
	Combining the cases above, we get the desired results with $b := \max\left(1-\frac{\delta_\beta}{c_gd_\constraintX},1-\frac{\delta_\beta}{\mu}\right)<1$. 
\end{proof}

\subsection{Proof of Theorem \ref{thm}} \label{sec:proofC}

Before we can prove Theorem~\ref{thm}, we first need the following auxiliary result.

\begin{lem} \label{lem:aux}
	Let $\{a_k\}_{k=0}^{M}$ be any sequence that satisfies $a_k\in[0,1]$ for all $M\in\mathbb{N}$ and all $k \in \mathbb{N}_{[0,M]}$, and let $\epsilon\in(0,1]$. For any $c\in(0,1]$, $\prod_{k=0}^{M} (1-a_k) \leq c$ implies $\prod_{k=0}^{M} (1-a_k\epsilon) \leq 1-(1-c)\epsilon$.
\end{lem}

\begin{proof}
	Define $m(c) := 1-(1-c)\epsilon$. We prove the result by induction on $M$. Note that the result is trivially true for $M=0$ since $1-a_0 \leq c$ implies $a_0\geq 1-c$ and, thus, $1-a_0\epsilon \leq 1-(1-c)\epsilon = m(c)$.
	
	In the following, assume that the result is true for some $M-1\in\mathbb{N}$, i.e., $\prod_{k=0}^{M-1}(1-a_k)\leq c$ implies $\prod_{k=0}^{M-1}(1-a_k\epsilon) \leq m(c)$. Define $d(a_0,\dots,a_{M-1}) := \prod_{k=0}^{M-1} (1-a_k) \leq 1$, where we omit the arguments of $d(a_0,\dots,a_{M-1})$ in the remainder of this proof. First, assume $d< c$. Then, we obtain $\prod_{k=0}^M (1-a_k\epsilon) \leq (1-a_M\epsilon) m(d) < m(d) \leq m(c)$. Second, let $d \geq c$. Then, $\prod_{k=0}^{M}(1-a_k) = (1-a_{M}) d \leq c$ implies $a_{M}\geq 1-\frac{c}{d}$. Furthermore, we get $\prod_{k=0}^{M} (1-a_k\epsilon) = (1-a_{M}\epsilon) \cdot \prod_{k=0}^{M-1} (1-a_k\epsilon) \leq (1-(1-\frac{c}{d})\epsilon) m(d) =: f(d)$. It is easy to see that the function $f:[c,1]\mapsto\mathbb{R}$ is continuously differentiable on an open set that contains its domain, that $\frac{\partial f}{\partial d}(d)=0$ if and only if $d=\sqrt{c}\in(c,1]$, and that $\frac{\partial^2 f}{\partial d^2}>0$ on its domain, i.e., $f(d)$ is strictly convex. Thus, $f(d)\leq f(1)=f(c)=m(c)$ for all $d\in[c,1]$, which concludes the proof by induction.
%
\end{proof}

Next, we prove Theorem~\ref{thm}.

\begin{proof}
Constraint satisfaction and existence of a feasible solution to~\eqref{algo:additional_input}-\eqref{algo:beta} follows from Lemma~\ref{lem:feasibility}. First, $\gamma\in\left(0,\frac{2}{\alpha_K+l_K}\right]$ together with Assumption~\ref{ass:cost_function} implies the following convergence rate of gradient descent in~\eqref{algo:OGD}
	\begin{equation}
		\norm{\estz_t - \zeta_{t-1}} \leq \kappa \norm{ \predz_{t} - \zeta_{t-1}}, \label{eq:GD_contr}
	\end{equation}
	where $\kappa = 1-\gamma\alpha_K \in [0,1)$ (compare, e.g., \cite[Theorem 2.2.14]{Nesterov18}), for any $t\in\mathbb{N}_{\geq1}$. Next, by the definition of dynamic regret and Lipschitz continuity in Assumption~\ref{ass:cost_function} we get
	\begin{align*}
		\regret &\refeq{\eqref{eq:def_regret}} \sum_{t=0}^T L_t(x_t,u_t) - L_t(\theta_t,\eta_t+K\theta_t) \\
		&\leq G \sum_{t=0}^T \norm{\begin{bmatrix} x_{t} \\ u_t \end{bmatrix} - \begin{bmatrix} \theta_t \\ \eta_t+K\theta_t \end{bmatrix}} \\
		&\refeq{\eqref{algo:control_input}} G\sum_{t=0}^T \norm{ \begin{bmatrix} \measx_t-\theta_t \\ T_1 \predu_t - \eta_t + K(\measx_t-\theta_t) \end{bmatrix} - \begin{bmatrix} v_t \\ \zero_n\end{bmatrix}} \\
		&\leq \tilde C_0 + G(\norm{K}+1) \sum_{t=\mu+1}^T \norm{\begin{bmatrix} \measx_t - \theta_t \\ T_{1} \predu_t - \eta_t \end{bmatrix}} + G\sum_{t=0}^T \norm{v_t},
	\end{align*}
	where $\tilde C_0 := G \sum_{t=0}^{\mu} \norm{ \begin{bmatrix} \measx_t-\theta_t \\ K\measx_t+T_1 \predu_t - (K\theta_t + \eta_t) \end{bmatrix} }$. Note that $\tilde C_0 \leq G(\mu+1) (d_\constraintX+d_\constraintU+r_{\mathcal V})$ holds by Lemma~\ref{lem:feasibility} and $\zeta_t\in\stronglyfeasibleSS$, where $r_{\mathcal V} := \max_{v\in\mathcal V} \norm{v}$. Then, defining $c_{GK}:=G(\norm{K}+1)$ and using the triangle inequality we get
	\begin{equation}
		\begin{split}
			\regret&\leq \tilde{C}_0 + c_{GK} \underbrace{\sum_{t=1}^{T-\mu} \norm{\begin{bmatrix} \measx_{t+\mu} \\ T_1 \predu_{t+\mu} \end{bmatrix} - \predz_t}}_\text{Part III} + G\sum_{t=0}^T \norm{v_t} \\ &\quad +c_{GK} \underbrace{\sum_{t=1}^{T-\mu}\norm{\predz_t - \zeta_t}}_\text{Part II}  + c_{GK} \underbrace{\sum_{t=1}^{T-\mu} \norm{\zeta_{t+\mu} - \zeta_t}}_\text{Part I}.
		\end{split} \label{eq:regret_bound_1}
	\end{equation}
	We proceed to bound the three sums in \eqref{eq:regret_bound_1} separately.
	
	\noindent\underline{\textit{Part I:}} We have
	\begin{align}
		&\sum_{t=1}^{T-\mu} \norm{\zeta_{t+\mu} - \zeta_{t}} = \sum_{t=1}^{T-\mu} \norm{\sum_{k=t+1}^{t+\mu} \zeta_k - \zeta_{k-1}} \nonumber \\
		\leq &\sum_{t=1}^{T-\mu}\sum_{k=t+1}^{t+\mu} \norm{\zeta_k - \zeta_{k-1}}
		\leq \mu \sum_{t=1}^T \norm{\zeta_t - \zeta_{t-1}}. \label{eq:regret_bound_part_1}
	\end{align}
	
	\noindent\underline{\textit{Part II:}} Using \eqref{eq:pred_ss_error_rec}, we get for any $t\in\mathbb{N}_{\geq2}$
	\begin{align*}
		&\norm{\predz_t-\ssz_{t-1}} \refeq{\eqref{algo:def_predz}} \norm{\begin{bmatrix} \predx_t \\ \ssinput_{t-1} \end{bmatrix} - \begin{bmatrix} \ssstate_{t-1} \\ \ssinput_{t-1} \end{bmatrix} } = \norm{\predx_t-\ssstate_{t-1}} \\
		\refleq{\eqref{eq:pred_ss_error_rec}} &(1-\beta_{t-1}) \norm{A_K(\predx_{t-1}-\ssstate_{t-2})} + \norm{A_K^\mu}\norm{\noise_t} \\
		\refleq{\eqref{eq:pred_ss_error_rec}} &\norm{A_K^{t-1}\left( \predx_{1}-\ssstate_{0} \right) } \prod_{j=0}^{t-2} (1-\beta_{1+j}) \\
		&\quad + \sum_{k=2}^{t} \left( \prod_{j=0}^{t-k-1} (1-\beta_{k+j}) \right) \norm{A_K^{t+\mu-k}} \norm{\noise_{k}} \\
		\refleq{\eqref{eq:predu_init}}&\sum_{k=1}^{t} \left( \prod_{j=0}^{t-k-1} (1-\beta_{k+j}) \right) \norm{A_K^{t+\mu-k}} \norm{\noise_{k}}.
	\end{align*}
	Additionally, we obtain the same result for $t=1$ by~\eqref{eq:predu_init}. Next, recall the constants $c_A\geq1$ and $\phi\in[0,1)$ such that $\norm{A_K^t}\leq c_A\phi^t\leq c_A$ holds for any $t\in\mathbb{N}$. Then, for any $\tau\in\mathbb{N}_{\geq1}$, summing over the above inequality yields
	\begin{align*}
		&\sum_{t=1}^\tau \norm{\predz_t-\ssz_{t-1}} \leq \sum_{t=1}^\tau \sum_{k=1}^{t} \left( \prod_{j=0}^{t-k-1} (1-\beta_{k+j}) \right) c_A \norm{\noise_{k}} \\
		\leq &c_A\sum_{t=1}^\tau \norm{\noise_t} \left( \sum_{k=t}^{\tau} \prod_{j=0}^{\tau-k-1} (1-\beta_{t+j}) \right) \\
		= &c_A\sum_{t=1}^\tau \norm{\noise_t} \left( 1+\sum_{k=0}^{\tau-t-1} \prod_{j=0}^{\tau-k-t-1} (1-\beta_{t+j}) \right),
	\end{align*}
	where we can use Lemma~\ref{lem:mincontr} to get
	\begin{align}
		\sum_{t=1}^\tau \norm{\predz_t-\ssz_{t-1}} &\leq c_A \sum_{t=1}^\tau \norm{\noise_t} \left( 1+ (\mu+1) \sum_{k=0}^{\lceil \frac{\tau-t}{\mu+1} \rceil} b^k \right) \nonumber \\
		&\leq c_A m_\beta \sum_{t=1}^\tau \norm{\noise_t}, \label{eq:predz-ssz}
	\end{align}
	where $m_\beta = 1+\frac{1+\mu}{1-b}$. Next, for any $t\in\mathbb{N}_{\geq1}$, we have
	\begin{align}
		&\norm{\ssz_t - \zeta_{t}} \leq \norm{\ssz_t - \zeta_{t-1}} + \norm{\zeta_{t}-\zeta_{t-1}}\nonumber \\
		\refleq{\eqref{eq:zs_convex_comb}} &(1-\beta_{t})\norm{ \ssz_{t-1} - \zeta_{t-1} }+ \beta_{t} \norm{\estz_{t} - \zeta_{t-1}} + \norm{\zeta_{t} - \zeta_{t-1}} \nonumber \\
		\refleq{\eqref{eq:GD_contr}} &(1-\beta_{t})\norm{ \ssz_{t-1} - \zeta_{t-1} } + \beta_{t} \kappa \norm{\predz_{t} - \zeta_{t-1}} + \norm{\zeta_{t} - \zeta_{t-1}} \nonumber \\
		\leq &~\tilde{\beta}_t\norm{\ssz_{t-1} - \zeta_{t-1}} + \beta_{t} \kappa \norm{\predz_{t}-\ssz_{t-1}} + \norm{\zeta_{t} - \zeta_{t-1}} \nonumber \\
		\leq &~\tilde{\beta}_{t}\norm{ \ssz_{t-1} - \zeta_{t-1} } + \kappa \norm{\predz_{t}-\ssz_{t-1}} + \norm{\zeta_{t} - \zeta_{t-1}}, \label{eq:ssz_rec}
	\end{align}
	where $\tilde{\beta}_t := 1-\beta_t(1-\kappa)$. Applying \eqref{eq:ssz_rec} recursively yields%
	\begin{align*}
		&\norm{\ssz_t - \zeta_{t}} \leq \left(\prod_{j=0}^{t-1} \tilde{\beta}_{1+j} \right) \norm{\ssz_{0} - \zeta_{0}} \\
		&\quad+ \sum_{k=1}^{t} \left( \prod_{j=0}^{t-k-1} \tilde{\beta}_{k+j+1} \right) \left(\kappa \norm{\predz_{k} - \ssz_{k-1}} + \norm{\zeta_k-\zeta_{k-1}} \right).
	\end{align*}
	Thus, for any $\tau \in \mathbb{N}_{\geq1}$, summing over the above inequality, applying Lemma~\ref{lem:mincontr}, Lemma~\ref{lem:aux} with $c=b>0$ and $\epsilon=1-\kappa$, and defining $c_{\kappa} := 1-(1-b)(1-\kappa)\in[0,1)$ leads to
	\begin{align}
		&\sum_{t=1}^{\tau} \norm{\ssz_t - \zeta_{t}} \leq \norm{\ssz_0-\zeta_0} \sum_{t=1}^{\tau} \prod_{j=0}^{t-1}\tilde{\beta}_{1+j} \nonumber \\
		&+ \sum_{t=1}^\tau\sum_{k=1}^{t}\left(\prod_{j=0}^{t-k-1} \tilde{\beta}_{k+j+1} \right) \left(\kappa \norm{\predz_{k} - \ssz_{k-1}} + \norm{\zeta_k-\zeta_{k-1}} \right) \nonumber \\
		&\leq \norm{\ssz_0-\zeta_0} (\mu+1)\sum_{k=0}^{\lceil \frac{\tau}{\mu+1}\rceil} c_\kappa^k \nonumber \\
		&\quad + \sum_{t=1}^\tau \norm{\zeta_t-\zeta_{t-1}} \left( 1+(\mu+1) \sum_{k=0}^{\lceil\frac{\tau-t}{\mu+1}\rceil} c_\kappa^k \right) \nonumber \\
		&\quad+\kappa \sum_{t=1}^\tau \norm{\predz_t-\ssz_{t-1}} \left( 1+(\mu+1) \sum_{k=0}^{\lceil\frac{\tau-t}{\mu+1}\rceil} c_\kappa^k \right) \nonumber \\
		&\leq (m_\kappa-1) \norm{\ssz_0-\zeta_0} + \kappa m_\kappa \sum_{t=1}^\tau \norm{\predz_t-\ssz_{t-1}} \nonumber \\
		&\quad + m_\kappa \sum_{t=1}^\tau \norm{\zeta_t-\zeta_{t-1}} \nonumber \\
		\begin{split} \label{eq:ssz-zeta}
		&\refleq{\eqref{eq:predz-ssz}} (m_\kappa-1) \norm{\ssz_0-\zeta_0} + c_A\kappa m_\beta m_\kappa \sum_{t=1}^\tau \norm{\noise_t} \\
		&\quad + m_\kappa \sum_{t=1}^\tau \norm{\zeta_t-\zeta_{t-1}}
		\end{split}
	\end{align}
	where $m_\kappa = 1+\frac{\mu+1}{1-c_\kappa}$. Finally, for any $\tau\in\mathbb{N}_{\geq1}$, we get the desired result
	\begin{align}
		&\sum_{t=1}^{\tau}\norm{\predz_t - \zeta_t} \leq \sum_{t=1}^\tau  \norm{\predz_t-\ssz_{t-1}} + \sum_{t=0}^{\tau-1}\norm{\ssz_{t} - \zeta_{t}} \nonumber \\
		&\qquad+ \sum_{t=1}^\tau\norm{\zeta_t - \zeta_{t-1}} \nonumber \\
		\begin{split} \label{eq:regret_bound_part_2}
		&\refleq{\eqref{eq:predz-ssz},\eqref{eq:ssz-zeta}} c_Am_\beta(1+\kappa m_\kappa) \sum_{t=1}^\tau \norm{\noise_t} + m_\kappa \norm{\ssz_0-\zeta_0} \\
		&\qquad + \left(m_\kappa+1\right) \sum_{t=1}^\tau \norm{\zeta_t - \zeta_{t-1}}. \end{split}
	\end{align}
	
	\noindent\underline{\textit{Part III:}}
	Note that \eqref{eq:regret_bound_part_2} implies
	\begin{align}
		&\sum_{t=1}^\tau \norm{\predz_t - \estz_t} \leq \sum_{t=1}^\tau \norm{ \predz_t - \zeta_{t-1}} + \sum_{t=0}^\tau \norm{\estz_t-\zeta_{t-1}} \nonumber \\
		\refleq{\eqref{eq:GD_contr}} &(1+\kappa) \sum_{t=1}^\tau \norm{\predz_t-\zeta_{t}} + (1+\kappa) \sum_{t=1}^\tau \norm{\zeta_t - \zeta_{t-1}} \nonumber \\
		\begin{split}\label{eq:zs-hatz}
			\refleq{\eqref{eq:regret_bound_part_2}} & (1+\kappa)m_\kappa \norm{\ssz_0-\zeta_0} + (1+\kappa)(m_\kappa+2) \sum_{t=1}^{\tau} \norm{\zeta_t - \zeta_{t-1}} \\
			&\quad+ (1+\kappa)c_Am_\beta(1+\kappa m_\kappa)  \sum_{t=1}^{\tau} \norm{\noise_t},
		\end{split}
	\end{align}
	and, using~\eqref{eq:zs-hatz} with $\tau=T-\mu$,
	\begin{align}
		&\sum_{t=1}^{T-\mu} \norm{\ssinput_t - \ssinput_{t-1}} \refeq{\eqref{algo:us_t}} \sum_{t=1}^{T-\mu} \beta_t \norm{\estu_t - \ssinput_{t-1}} \refleq{\eqref{algo:def_predz}} \sum_{t=1}^{T-\mu} \norm{\estz_t-\predz_t} \nonumber \\
		\begin{split}\label{eq:diff_us_t}
			&\refleq{\eqref{eq:zs-hatz}} (1+\kappa)m_\kappa \norm{\ssz_0-\zeta_0} + (1+\kappa)(m_\kappa+2) \sum_{t=1}^{T-\mu} \norm{\zeta_t - \zeta_{t-1}} \\
			&\quad+ (1+\kappa)c_Am_\beta(1+\kappa m_\kappa)  \sum_{t=1}^{T-\mu} \norm{\noise_t}.
		\end{split} 
	\end{align}	
	Moreover, we get for $k\in\mathbb{N}_{[0,\mu]}$ by repeatedly using~\eqref{algo:predicted_input}
	\begin{align}
		&T_1\predu_{t+k} \refeq{\eqref{algo:predicted_input}} T_1 \begin{bmatrix} \sigma \predu_{t+k-1} \\ \ssinput_{t+k-1} \end{bmatrix} + \beta_{t+k} T_1 g_{t+k} \nonumber \\
		\refeq{\eqref{algo:predicted_input}} ~&T_{2} \begin{bmatrix} \sigma \predu_{t+k-2} \\ \ssinput_{t+k-2} \end{bmatrix} + \sum_{j=k-1}^k \beta_{t+j} T_{j-k+1} g_{t+j} \nonumber \\
		\refeq{\eqref{algo:predicted_input}}~ &\begin{cases} 
			&T_{2+k} \predu_{t-1} + \sum_{j=0}^k \beta_{t+j} T_{k-j+1} g_{t+j} \\
			&\hspace{4cm}\text{if } k\in[0,\mu-2] \\
			&\ssinput_{t-1} + \sum_{j=0}^{\mu-1} \beta_{t+j} T_{\mu-j} g_{t+j} \\
			&\hspace{4cm}\text{if } k = \mu-1 \\ 
			&\ssinput_t + \sum_{j=1}^\mu \beta_{t+j} T_{\mu-j+1} g_{t+j} \\
			&\hspace{4cm}\text{if } k=\mu \end{cases}\label{eq:input_t+mu}
	\end{align}
	Furthermore, from applying \eqref{eq:meas_state_dynamics} and \eqref{algo:control_input} repeatedly, we get 
	\[
		\tilde x_{t+\mu} = A_K^\mu \measx_t + \controlMatrix \begin{bmatrix} T_1 \predu_{t} \\ \vdots \\ T_1\predu_{t+\mu-1} \end{bmatrix} + \sum_{k=0}^{\mu-1} A_K^k \noise_{t+\mu-k}
	\]
	for any $t\in\mathbb{N}$. Finally, using this results it follows that
	\begin{align*}
		&\norm{\predz_t-\begin{bmatrix} \measx_{t+\mu} \\ T_1 \predu_{t+\mu} \end{bmatrix}} \refleq{\eqref{algo:prediction}} \norm{\begin{bmatrix} \controlMatrix \left( \begin{bmatrix} \sigma \predu_{t-1} \\ \ssinput_{t-1} \end{bmatrix} - \begin{bmatrix} T_1 \predu_t \\ \vdots \\ T_1\predu_{t+\mu-1} \end{bmatrix} \right) \\ \ssinput_{t-1} - T_1 \predu_{t+\mu} \end{bmatrix}} \\
		&\qquad+\norm{\sum_{k=0}^{\mu-1} A_K^k \noise_{t+\mu-k}} 
	\end{align*}
	\begin{align*}
		&\refleq{\eqref{eq:input_t+mu}} \norm{ \begin{bmatrix} -\controlMatrix \begin{bmatrix}\beta_t T_1 g_t \\ \vdots \\ \sum_{j=0}^{\mu-1} \beta_{t+j} T_{\mu-j} g_{t+j} \end{bmatrix} \\ \ssinput_{t-1} - \ssinput_t - \sum_{j=1}^{\mu} \beta_{t+j} T_{\mu-j+1} g_{t+j} \end{bmatrix} } \\
		&\qquad+ \sum_{k=0}^{\mu-1} \norm{A_K^k} \norm{\noise_{t+\mu-k}} \\
		&\leq \norm{\ssinput_{t-1} - \ssinput_t} + \norm{\controlMatrix} \sum_{k=0}^{\mu-1} \norm{\sum_{j=0}^k \beta_{t+j} T_{k-j+1} g_{t+j} } \\
		&\qquad +\sum_{j=1}^\mu \beta_{t+j} \norm{T_{\mu-j+1} g_{t+j}} + \sum_{k=0}^{\mu-1} \norm{A_K^k} \norm{\noise_{t+\mu-k}}. 
	\end{align*}
	holds for any $t\in\mathbb{N}$. From here, we use $\beta_t\leq1$ for all $t\in\mathbb{N}$, $\norm{T_i}=1$ for all $i\in\mathbb{N}_{[1,\mu]}$, and~\eqref{algo:additional_input} to obtain
	\begin{align*}
		&\norm{\predz_t-\begin{bmatrix} \measx_{t+\mu} \\ T_1 \predu_{t+\mu} \end{bmatrix}}\refleq{\eqref{algo:additional_input}} \norm{\ssinput_{t}-\ssinput_{t-1}} + c_g \sum_{j=1}^{\mu} \norm{\estx_{t+j}-\predx_{t+j}} \\
		&\quad+ \norm{\controlMatrix} c_g \sum_{k=0}^{\mu-1} \sum_{j=0}^k \norm{\estx_{t+j}-\predx_{t+j}} +\sum_{k=0}^{\mu-1} \norm{A_K^k} \norm{\noise_{t+\mu-k}}.
	\end{align*}
	Summing over the above inequality, we get
	\begin{align*}
		&\sum_{t=1}^{T-\mu} \norm{\predz_t-\begin{bmatrix} \measx_{t+\mu} \\ T_1 \predu_{t+\mu}\end{bmatrix}} \leq \sum_{t=1}^{T-\mu}\norm{\ssinput_{t-1}-\ssinput_t}  \\
		&\qquad + c_g \mu \sum_{t=1}^{T} \norm{\estx_{t}-\predx_{t}}  \\
		&\qquad+ \norm{\controlMatrix} c_g \mu \sum_{t=1}^{T-\mu} \sum_{j=0}^{\mu-1} \norm{\estx_{t+j}-\predx_{t+j}}  \\
		&\qquad + \sum_{t=1}^{T} \norm{\noise_{t}} \left( \sum_{k=0}^{\mu-1} c_A\phi^k \right)   \\
		&\leq \sum_{t=1}^{T-\mu} \norm{\ssinput_{t}{-}\ssinput_{t-1}} + (c_z{-}1) \sum_{t=1}^T \norm{\estz_t{-}\predz_t}  + \frac{c_A}{1{-}\phi} \sum_{t=1}^T \norm{\noise_t},
	\end{align*}
	where $c_z=c_g\mu\left(1+\norm{\controlMatrix}\mu\right)+1$. Finally, inserting~\eqref{eq:zs-hatz} with $\tau=T$ and~\eqref{eq:diff_us_t} yields
	\begin{align}
		\begin{split}\label{eq:regret_bound_part_3}
		&\sum_{t=1}^{T-\mu} \norm{\predz_t-\begin{bmatrix} \measx_{t+\mu} \\ T_1 \predu_{t+\mu}\end{bmatrix}} \leq c_z(1+\kappa)m_\kappa\norm{\ssz_0-\zeta_0}   \\
		&\mkern0mu+ c_z (1+\kappa)(m_\kappa+2) \sum_{t=1}^T \norm{\zeta_t-\zeta_{t-1}} + c_{\noise} \sum_{t=1}^T \norm{\noise_t},
		\end{split}
	\end{align}	
	where $c_{\noise} = c_A\frac{c_z(1+\kappa)m_\beta(1+\kappa m_\kappa)(1-\phi)+1}{1-\phi}$. Noting that 
	\[
		\sum_{t=1}^{T} \norm{\noise_t} \leq \sum_{t=0}^{T-1} \norm{w_t} + \left(1+\norm{A}\right)\sum_{t=0}^T \norm{v_t}
	\]
	and $\norm{\ssz_0-\zeta_0}\leq d_\constraintZ$, the result then follows from inserting \eqref{eq:regret_bound_part_1}, \eqref{eq:regret_bound_part_2} with $\tau = T-\mu$, and \eqref{eq:regret_bound_part_3} into \eqref{eq:regret_bound_1}.
\end{proof}

\section*{References}

\bibliography{bib.bib}

\begin{thebibliography}{10}
\providecommand{\url}[1]{#1}
\csname url@samestyle\endcsname
\providecommand{\newblock}{\relax}
\providecommand{\bibinfo}[2]{#2}
\providecommand{\BIBentrySTDinterwordspacing}{\spaceskip=0pt\relax}
\providecommand{\BIBentryALTinterwordstretchfactor}{4}
\providecommand{\BIBentryALTinterwordspacing}{\spaceskip=\fontdimen2\font plus
\BIBentryALTinterwordstretchfactor\fontdimen3\font minus
  \fontdimen4\font\relax}
\providecommand{\BIBforeignlanguage}[2]{{%
\expandafter\ifx\csname l@#1\endcsname\relax
\typeout{** WARNING: IEEEtran.bst: No hyphenation pattern has been}%
\typeout{** loaded for the language `#1'. Using the pattern for}%
\typeout{** the default language instead.}%
\else
\language=\csname l@#1\endcsname
\fi
#2}}
\providecommand{\BIBdecl}{\relax}
\BIBdecl

\bibitem{Shalev-Shwartz2012}
S.~Shalev-Shwartz, ``Online learning and online convex optimization,''
  \emph{Foundations and Trends{\textregistered} in Machine Learning}, vol.~4,
  no.~2, pp. 107--194, 2012.

\bibitem{Hazan2016}
E.~Hazan, ``Introduction to online convex optimization,'' \emph{Foundations and
  Trends{\textregistered} in Optimization}, vol.~2, no. 3-4, pp. 157--325,
  2016.

\bibitem{Tang2017}
Y.~Tang, K.~Dvijotham, and S.~Low, ``Real-time optimal power flow,'' \emph{IEEE
  Transactions on Smart Grid}, vol.~8, no.~6, pp. 2963--2973, 2017.

\bibitem{Zheng2020}
T.~Zheng, J.~Simpson-Porco, and E.~Mallada, ``Implicit trajectory planning for
  feedback linearizable systems: A time-varying optimization approach,'' in
  \emph{Proc. 2020 American Control Conference}, 2020, pp. 4677--4682.

\bibitem{Li2019}
Y.~Li, X.~Chen, and N.~Li, ``Online optimal control with linear dynamics and
  predictions: Algorithms and regret analysis,'' in \emph{Advances in Neural
  Information Processing Systems}, 2019, pp. 14\,858 -- 14\,870.

\bibitem{Agarwal2019}
N.~Agarwal, B.~Bullins, E.~Hazan, S.~Kakade, and K.~Singh, ``Online control
  with adversarial disturbances,'' in \emph{Proc. 36th International Conference
  on Machine Learning}, vol.~97, 2019, pp. 111--119.

\bibitem{Shi2020}
G.~Shi, Y.~Lin, S.-J. Chung, Y.~Yue, and A.~Wierman, ``Online optimization with
  memory and competitive control,'' in \emph{Advances in Neural Information
  Processing Systems}, vol.~33.\hskip 1em plus 0.5em minus 0.4em\relax Curran
  Associates, Inc., 2020, pp. 20\,636--20\,647.

\bibitem{Hazan2022}
E.~Hazan and K.~Singh, ``Introduction to online nonstochastic control,'' 2022,
  available online at arXiv:2211.09619.

\bibitem{Nonhoff2022a}
M.~Nonhoff and M.~A. Müller, ``Online convex optimization for data-driven
  control of dynamical systems,'' \emph{IEEE Open Journal of Control Systems},
  vol.~1, pp. 180--193, 2022.

\bibitem{Lin2023}
Y.~Lin, J.~A. Preiss, E.~T. Anand, Y.~Li, Y.~Yue, and A.~Wierman, ``Online
  adaptive policy selection in time-varying systems: No-regret via contractive
  perturbations,'' in \emph{Thirty-seventh Conference on Neural Information
  Processing Systems}, 2023.

\bibitem{Karapetyan2023}
A.~Karapetyan, D.~Bolliger, A.~Tsiamis, E.~C. Balta, and J.~Lygeros, ``Online
  linear quadratic tracking with regret guarantees,'' \emph{IEEE Control
  Systems Letters}, vol.~7, pp. 3950--3955, 2023.

\bibitem{Didier2022}
A.~Didier, J.~Sieber, and M.~N. Zeilinger, ``A system level approach to regret
  optimal control,'' \emph{IEEE Control Systems Letters}, vol.~6, pp.
  2792--2797, 2022.

\bibitem{Goel2021}
G.~Goel and B.~Hassibi, ``Regret-optimal measurement-feedback control,'' in
  \emph{Proc. 3rd Conference on Learning for Dynamics and Control}, vol. 144,
  2021, pp. 1270--1280.

\bibitem{Martin2022}
A.~Martin, L.~Furieri, F.~D\"orfler, J.~Lygeros, and G.~Ferrari-Trecate, ``Safe
  control with minimal regret,'' in \emph{Proc. 4th Annual Learning for
  Dynamics and Control Conference}, vol. 168, 2022, pp. 726--738.

\bibitem{Martin2023}
A.~Martin, L.~Furieri, F.~Dörfler, J.~Lygeros, and G.~Ferrari-Trecate, ``On
  the guarantees of minimizing regret in receding horizon,'' \emph{IEEE
  Transactions on Automatic Control}, vol.~70, no.~3, pp. 1547--1562, 2025.

\bibitem{Zhou23CDC}
H.~Zhou and V.~Tzoumas, ``Safe control of partially-observed linear
  time-varying systems with minimal worst-case dynamic regret,'' in \emph{2023
  62nd IEEE Conference on Decision and Control (CDC)}, 2023, pp. 8781--8787.

\bibitem{Gharbi2021}
M.~Gharbi, B.~Gharesifard, and C.~Ebenbauer, ``Anytime proximity moving horizon
  estimation: Stability and regret for nonlinear systems,'' in \emph{Proc. 2021
  IEEE Conference on Decision and Control}, 2021, pp. 728--735.

\bibitem{Nonhoff2023a}
M.~Nonhoff and M.~A. Müller, ``On the relation between dynamic regret and
  closed-loop stability,'' \emph{Systems \& Control Letters}, vol. 177, p.
  105532, 2023.

\bibitem{Karapetyan2022}
A.~Karapetyan, A.~Tsiamis, E.~C. Balta, A.~Iannelli, and J.~Lygeros,
  ``Implications of regret on stability of linear dynamical systems,''
  \emph{IFAC-PapersOnLine}, vol.~56, no.~2, pp. 2583--2588, 2023.

\bibitem{Nonhoff2021}
M.~Nonhoff and M.~A. Müller, ``An online convex optimization algorithm for
  controlling linear systems with state and input constraints,'' in \emph{Proc.
  2021 American Control Conference (ACC)}, 2021, pp. 2523--2528.

\bibitem{Li2021}
Y.~Li, S.~Das, and N.~Li, ``Online optimal control with affine constraints,''
  \emph{Proc. AAAI Conference on Artificial Intelligence}, vol.~35, no.~10, pp.
  8527--8537, 2021.

\bibitem{Nonhoff2024}
M.~Nonhoff, M.~T.~A. Torshan, and M.~A. Müller, ``Robust control of
  constrained linear systems using online convex optimization and a reference
  governor,'' in \emph{Proc. IEEE 63rd Conference on Decision and Control},
  2024, pp. 6553--6559.

\bibitem{Zhou23}
H.~Zhou, Y.~Song, and V.~Tzoumas, ``Safe non-stochastic control of
  control-affine systems: An online convex optimization approach,'' \emph{IEEE
  Robotics and Automation Letters}, vol.~8, no.~12, pp. 7873--7880, 2023.

\bibitem{Nonhoff2025}
M.~Nonhoff, J.~Köhler, and M.~A. Müller, ``Online convex optimization for
  constrained control of nonlinear systems,'' 2024, available online at
  arXiv:2412.00922.

\bibitem{Simonetto2020}
A.~Simonetto, E.~Dall'Anese, S.~Paternain, G.~Leus, and G.~B. Giannakis,
  ``Time-varying convex optimization: Time-structured algorithms and
  applications,'' \emph{Proceedings of the IEEE}, vol. 108, no.~11, pp.
  2032--2048, 2020.

\bibitem{Menta2018}
S.~Menta, A.~Hauswirth, S.~Bolognani, G.~Hug, and F.~Dörfler, ``Stability of
  dynamic feedback optimization with applications to power systems,'' in
  \emph{Proc. 2018 56th Annual Allerton Conference on Communication, Control,
  and Computing}, 2018, pp. 136--143.

\bibitem{Colombino2020}
M.~{Colombino}, E.~{Dall'Anese}, and A.~{Bernstein}, ``Online optimization as a
  feedback controller: Stability and tracking,'' \emph{IEEE Transactions on
  Control of Network Systems}, vol.~7, no.~1, pp. 422--432, 2020.

\bibitem{Cothren2022}
L.~Cothren, G.~Bianchin, and E.~Dall'Anese, ``Online optimization of dynamical
  systems with deep learning perception,'' \emph{IEEE Open Journal of Control
  Systems}, vol.~1, pp. 306--321, 2022.

\bibitem{Lawrence2021}
L.~S.~P. Lawrence, J.~W. Simpson-Porco, and E.~Mallada, ``Linear-convex optimal
  steady-state control,'' \emph{IEEE Transactions on Automatic Control},
  vol.~66, no.~11, pp. 5377--5384, 2021.

\bibitem{Bianchin2022}
G.~Bianchin, J.~Cortés, J.~I. Poveda, and E.~Dall’Anese, ``Time-varying
  optimization of {LTI} systems via projected primal-dual gradient flows,''
  \emph{IEEE Transactions on Control of Network Systems}, vol.~9, no.~1, pp.
  474--486, 2022.

\bibitem{Dean2021}
S.~Dean and B.~Recht, ``Certainty equivalent perception-based control,'' in
  \emph{Proc. 3rd Conference on Learning for Dynamics and Control}, ser. Proc.
  of Machine Learning Research, vol. 144.\hskip 1em plus 0.5em minus
  0.4em\relax PMLR, 2021, pp. 399--411.

\bibitem{Marchi2022}
M.~Marchi, J.~Bunton, B.~Gharesifard, and P.~Tabuada, ``Safety and stability
  guarantees for control loops with deep learning perception,'' \emph{IEEE
  Control Systems Letters}, vol.~6, pp. 1286--1291, 2022.

\bibitem{Schenato2014}
L.~Schenato, G.~Barchi, D.~Macii, R.~Arghandeh, K.~Poolla, and A.~Von~Meier,
  ``Bayesian linear state estimation using smart meters and pmus measurements
  in distribution grids,'' in \emph{Proc. 2014 IEEE International Conference on
  Smart Grid Communications}, 2014, pp. 572--577.

\bibitem{Picallo2018}
M.~Picallo, A.~Anta, B.~De~Schutter, and A.~Panosyan, ``A two-step distribution
  system state estimator with grid constraints and mixed measurements,'' in
  \emph{Proc. 2018 Power Systems Computation Conference}, 2018, pp. 1--7.

\bibitem{Rawlings2017}
J.~B. Rawlings, D.~Q. Mayne, and M.~Diehl, \emph{Model predictive control:
  theory, computation, and design}.\hskip 1em plus 0.5em minus 0.4em\relax
  Madison, Wisconsin, USA: Nob Hill Publishing, 2017.

\bibitem{Chisci2001}
L.~Chisci, J.~Rossiter, and G.~Zappa, ``Systems with persistent disturbances:
  predictive control with restricted constraints,'' \emph{Automatica}, vol.~37,
  no.~7, pp. 1019--1028, 2001.

\bibitem{Mueller2015}
M.~A. Müller, L.~Grüne, and F.~Allgöwer, ``On the role of dissipativity in
  economic model predictive control,'' \emph{IFAC-PapersOnLine}, vol.~48,
  no.~23, pp. 110--116, 2015.

\bibitem{Angeli2012}
D.~Angeli, R.~Amrit, and J.~B. Rawlings, ``On average performance and stability
  of economic model predictive control,'' \emph{IEEE Transactions on Automatic
  Control}, vol.~57, no.~7, pp. 1615--1626, 2012.

\bibitem{Rakovic2006}
S.~V. Rakovic and K.~I. Kouramas, ``The minimal robust positively invariant set
  for linear discrete time systems: Approximation methods and control
  applications,'' in \emph{Proc. 45th IEEE Conference on Decision and Control},
  2006, pp. 4562--4567.

\bibitem{Kolmanovsky95}
I.~Kolmanovsky and E.~Gilbert, ``Maximal output admissible sets for
  discrete-time systems with disturbance inputs,'' in \emph{Proc. 1995 American
  Control Conference}, vol.~3, 1995, pp. 1995--1999.

\bibitem{Koehler2020}
J.~Köhler, M.~A. Müller, and F.~Allgöwer, ``A nonlinear tracking model
  predictive control scheme for dynamic target signals,'' \emph{Automatica},
  vol. 118, p. 109030, 2020.

\bibitem{Zinkevich03}
M.~Zinkevich, ``Online convex programming and generalized infinitesimal
  gradient ascent,'' in \emph{Proc. 20th international conference on machine
  learning}, 2003, pp. 928--936.

\bibitem{Necoara2019}
I.~Necoara, Y.~Nesterov, and F.~Glinearu, ``Linear convergence of first order
  methods for non-strongly convex optimization,'' \emph{Mathematical
  Programming}, vol. 175, pp. 69--107, 2019.

\bibitem{Faulwasser18}
T.~Faulwasser, L.~Gr\"une, and M.~A. M\"uller, ``Economic nonlinear model
  predictive control,'' \emph{Foundations and
  Trends\textsuperscript{\textregistered} in Systems and Control}, vol.~5,
  no.~1, pp. 1--98, 2018.

\bibitem{Li2021a}
Y.~Li, G.~Qu, and N.~Li, ``Online optimization with predictions and switching
  costs: Fast algorithms and the fundamental limit,'' \emph{IEEE Transactions
  on Automatic Control}, vol.~66, no.~10, pp. 4761--4768, 2021.

\bibitem{Rakovic2005}
S.~Rakovic, E.~Kerrigan, K.~Kouramas, and D.~Mayne, ``Invariant approximations
  of the minimal robust positively invariant set,'' \emph{IEEE Transactions on
  Automatic Control}, vol.~50, no.~3, pp. 406--410, 2005.

\bibitem{MPT3}
M.~Herceg, M.~Kvasnica, C.~Jones, and M.~Morari, ``{Multi-Parametric Toolbox
  3.0},'' in \emph{Proc. 2013 European Control Conference}, 2013, pp. 502--510,
  \url{http://control.ee.ethz.ch/~mpt}.

\bibitem{casadi}
J.~A.~E. Andersson, J.~Gillis, G.~Horn, J.~B. Rawlings, and M.~Diehl,
  ``{CasADi} -- {A} software framework for nonlinear optimization and optimal
  control,'' \emph{Mathematical Programming Computation}, vol.~11, no.~1, pp.
  1--36, 2019.

\bibitem{Nesterov18}
Y.~Nesterov, \emph{Lectures on Convex Optimization}, 2nd~ed., ser. Springer
  Optimization and Its Applications.\hskip 1em plus 0.5em minus 0.4em\relax
  Cham, Switzerland: Springer, 2018, vol. 137.

\end{thebibliography}
\bibliographystyle{IEEEtran}

\begin{IEEEbiography}[{\includegraphics[width=1in,height=1.25in,clip,keepaspectratio]{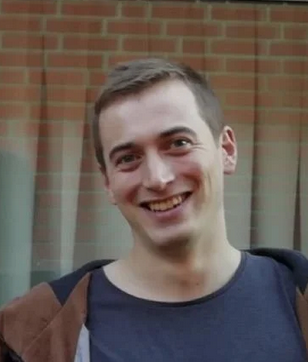}}]{Marko Nonhoff} (Member, IEEE) received his Master's degree in engineering cybernetics from the University of Stuttgart, Germany, in 2018, and a Ph.D. from the Leibniz University Hannover, Germany, in 2025. Since then, he is with the Leibniz University Hannover, Germany, as a postdoc. His research interests include learning-based control, optimal control, and online optimization.
\end{IEEEbiography}

\begin{IEEEbiography}[{\includegraphics[width=1in,height=1.25in,clip,keepaspectratio]{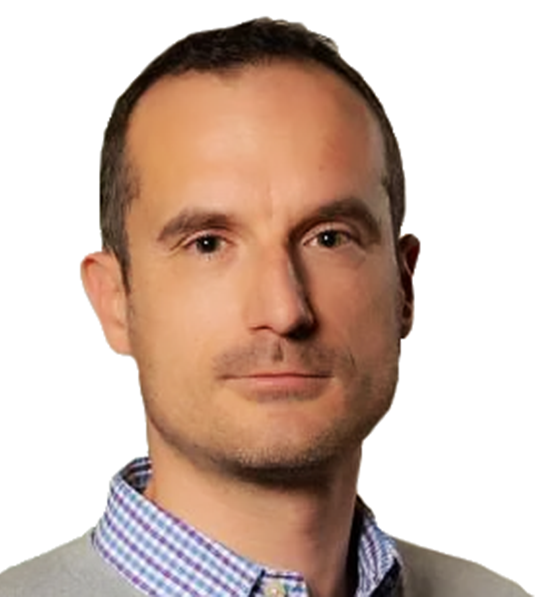}}]{Emiliano Dall'Anese} (Senior Member, IEEE) is an Associate Professor in the Department of Electrical and Computer Engineering at Boston University, where he is  also a faculty with the Division of Systems Engineering. He received  the Ph.D. in Information Engineering from the Department of Information Engineering, University of Padova, Italy, in 2011. He was with the University of Minnesota as a postdoc (2011-2014), the National Renewable Energy Laboratory as a senior researcher (2014-2018), and the Department of Electrical, Computer, and Energy Engineering at the University of Colorado Boulder as a faculty (2018-2024). 
	
His research interests span the areas of optimization, control, and learning; current applications include power systems and autonomous systems. He received the  National Science Foundation CAREER Award in 2020, the IEEE PES Prize Paper Award in 2021, and the IEEE Transactions on Control of Network Systems Best Paper Award in 2023.
\end{IEEEbiography}

\begin{IEEEbiography}[{\includegraphics[width=1in,height=1.25in,clip,keepaspectratio]{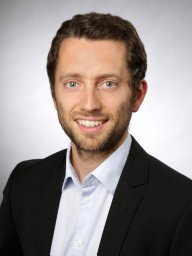}}]{Matthias A. M{\"u}ller} (Senior Member, IEEE) received a Diploma degree in engineering cybernetics from the University of Stuttgart, Germany, an M.Sc. in electrical and computer engineering from the University of Illinois at Urbana-Champaign, US (both in 2009), and a Ph.D. from the University of Stuttgart in 2014. Since 2019, he is Director of the Institute of Automatic Control and Full Professor at the Leibniz University Hannover, Germany. 
	
His research interests include nonlinear control and estimation, model predictive control, and data- and learning-based control, with application in different fields including biomedical engineering and robotics. He has received various awards for his work, including the 2015 European Systems \& Control PhD Thesis Award, the inaugural Brockett-Willems Outstanding Paper Award for the best paper published in Systems \& Control Letters in the period 2014-2018, an ERC starting grant in 2020, the IEEE CSS George S. Axelby Outstanding Paper Award 2022, and the Journal of Process Control Paper Award 2023. He serves/d as an associate editor for Automatica and as an editor of the International Journal of Robust and Nonlinear Control.
\end{IEEEbiography}

\end{document}